\documentclass{article}
\usepackage{graphicx} 

\usepackage{graphicx} 
\usepackage[utf8]{inputenc}
\usepackage[margin=2.5cm]{geometry}
\usepackage{amsmath}
\usepackage{amssymb}
\usepackage{cite}
\usepackage{authblk}
\usepackage{graphicx}
\usepackage{xcolor}
\usepackage{cancel}
\usepackage{ulem}
\definecolor{link}{rgb}{0.0, 0.0, 0.0 }
\usepackage[bookmarks = true,
			pdfstartview = FitH,
			colorlinks = true,
			urlcolor=link,
			citecolor=link,
			linkcolor=link,
			hyperfootnotes=false]{hyperref}

\begin{document}
\title{Normal-ordered perturbative expansion for matter systems interacting with continuous-mode quantum photon fields}

\author[1,2]{Liwen Ko\thanks{liwen.jko@berkeley.edu}}
\author[1,2]{Robert L. Cook\thanks{rlcook@berkeley.edu}}
\author[1,2]{K. Birgitta Whaley\thanks{whaley@berkeley.edu}}
\affil[1]{Department of Chemistry, University of California, Berkeley, CA 94720, USA}
\affil[2]{Kavli Energy Nanoscience Institute at Berkeley, Berkeley, CA 94720, USA}

\maketitle

\begin{abstract}
We develop a normal-ordered perturbative expansion method to compute the reduced dynamics of a matter system interacting with a general continuous-mode photon field state. In this expansion, all field correlation functions are automatically normal-ordered, while the free evolution is modified by spontaneous emission. Our normal-ordered expansion elucidates the physical meaning of non-normal-ordered correlation terms in the conventional expansion. Furthermore, under intermediate coupling strengths, the normal-ordered expansion converges significantly faster than the conventional expansion, allowing one to extend the applicability of perturbative expansion to much larger matter-light coupling strengths. Interestingly, given an $m$-photon Fock state input, the normal-ordered expansion truncates exactly at the $2m$-th order. For special classes of input states, we show the normal-ordered expansion is consistent with the corresponding master equations. We demonstrate the normal-ordered expansion numerically for a two-level system interacting with coherent state, $m$-photon Fock state, and cat state inputs.
\end{abstract}

\section{Introduction}
The interaction between non-classical light and atomic systems has been of interest recently due to its potential applications in quantum technologies \cite{kimble2008quantum, Combes_2017_review}. On the other hand, the use of non-classical light to probe molecules, nano-structures, and materials have also gained attention because it allows one to extract information about matter systems that would be difficult to extract using classical-like light \cite{Mukamel_Rev_Mod_Phys, Schlawin_2017_tutorial, lupton2021review, amgar2019higher, li2023single}. An important aspect of understanding the quantum light-matter interaction is to understand the reduced dynamics of the matter system. This is especially important for the coherent control of atom, molecules, or superconducting qubits \cite{schlawin2017theory,brumer1986control, beyvers2006optimal,warner2025coherent,gu2022wave,qi2021manipulation,mitra2022quantum,Baragiola_2012}.
\par
A common method to describe the time evolution of the reduced matter system is the master equation approach, which are differential equations for the system state $\rho_{\text{sys}}(t)$. Master equations have been developed for systems interacting with photon fields in coherent states \cite{wiseman_milburn_2009_book}, $m$-photon Fock state pulses \cite{Baragiola_2012}, Markovian thermal states \cite{Gardiner1985}, Markovian squeezed states \cite{Gardiner1985}, and pulsed squeezed states \cite{gross2022master}. Although these master equations provide near-exact and compact descriptions for the time evolution of $\rho_{\text{sys}}(t)$, master equations are applicable only for a small class of initial photon states, as enumerated above. More numerical open quantum system methods such as the hierarchical equations of motion (HEOM) \cite{Tanimura_2020} or the quasi-adiabatic propagator path integral (QuAPI) \cite{makri1998quantum} also suffer from the same problem because these methods typically require a Gaussian initial state for the bosonic bath. A general photon state (e.g., an $m$-photon Fock state pulse or a cat state) can be non-Gaussian, making these methods inapplicable. 
\par
An alternative method to treat more general input states is the perturbative approach, where one treats the light-matter coupling as the small parameter and expands the time evolution as a perturbative series \cite{Schlawin_2017_tutorial, Mukamel_Rev_Mod_Phys}. While master equations are differential equations, perturbative expansions take the form of integral expressions. 
The perturbative approach can be applied to an arbitrary initial photon field state. However, in practice, one usually needs to truncate the perturbative expansion to some finite order. Therefore, it is accurate only when the light-matter coupling strength is sufficiently weak. 
\par
In the perturbative expansion for the reduced system state $\rho_{\text{sys}}(t)$, the $n$-th order effect of the photon field on the system is encoded in the $n$-th order field correlation functions (e.g., a 4-th order correlation function can take the form $\langle a(t_1)a^\dagger(t_2)a(t_3)a^\dagger(t_4)\rangle $). In the conventional perturbative expansion, the field operators in the correlation functions follow a type of time-ordering. This means that the correlation functions are in general not normal-ordered. To evaluate these correlation functions, one typically re-expressed them as sums of normal-ordered correlation functions using the commutation relation or Wick's theorem. 
\par
In this paper, we derive a normal-ordered perturbative expansion for $\rho_{\text{sys}}(t)$, in which all field correlation functions are automatically normal-ordered. The normal-ordered expansion relies on the assumption that the initial total state is a product of the initial system state and the initial field state, i.e., $\rho_{\text{tot}}(0)=\rho_{\text{sys}}(0)\otimes\rho_{\text{field}}(0)$.
We show mathematically that the non-normal-ordered-ness in the conventional expansion can be captured by a spontaneous emission term in our normal-ordered expansion. Having only normal-ordered correlation functions greatly simplifies the expressions for the perturbative expansion. Furthermore, we find that the normal-ordered expansion is significantly more accurate numerically compared to the conventional expansion method at the same expansion order, allowing one to extend the applicability of perturbative expansions to much larger matter-light coupling strengths. We note that Ref. \cite{gough2006quantum} has pointed out the connection between time-ordered exponential operators and normal-ordered exponential operators in the language of quantum stochastic calculus. In this work, we work in the language of ordinary calculus, and we establish a connection between time-ordering and normal-ordering for the time-evolution superoperator of the reduced system state.
\par
In the remaining parts of the paper, we will show the derivation of the normal-ordered perturbative expansion. Then we draw connections between the normal-ordered expansion and various master equations. Specifically, we show that master equations for coherent state, $m$-photon Fock state, and Markovian Gaussian state inputs can all be derived exactly from the infinite-order normal-ordered perturbative expansion. Finally, we provide numerical examples for coherent state, $m$-photon Fock state, and cat state inputs, showing that the normal-ordered expansion is significantly more accurate than the conventional expansion. We note that master equations for a cat state input have not been developed (to our knowledge), and semi-classical approaches fail due to the neglect of the interference between two coherent states. Our normal-ordered expansion thus provides an accurate method to compute the reduced system dynamics under the excitation of a a cat state for a wide range of light-matter coupling strengths.

\section{System+field Hamiltonian}
We write the total Hamiltonian as
\begin{equation}
    H_{\text{sys+field}} = H_{\text{sys}} + H_{\text{field}} + H_{\text{coup}}.
\end{equation}
$H_{\text{sys}}$ is the matter Hamiltonian. $H_{\text{field}}$ is the photon field Hamiltonian, taking the form
\begin{equation}
    H_{\text{field}} = \sum_l \int^\infty_0 d\omega\, \omega a^\dagger_l(\omega)a_l(\omega),
\end{equation}
where we set $\hbar=1$.
$l$ indexes the photon field modes. $a_l(\omega)$ is the photon annihilation operator of mode $l$ at frequency $\omega$. The system-field coupling term is 
\begin{equation}
    H_{\text{coup}} = \sum_l \int^\infty_0 \frac{d\omega}{\sqrt{2\pi}}\,-ia_l(\omega)L^\dagger_l + ia^\dagger_l(\omega)L_l,
\end{equation}
taking the form of dipole -- electric field coupling. $L_l$ is the matter de-excitation part of the dipole operator that is coupled to the field mode $l$. We have taken the rotating wave approximation to conserve the total (matter and field) excitation number in $H_{\text{coup}}$. Using a narrow-band approximation to extend the $d\omega$ integration range from $(0, \infty)$ to $(-\infty, \infty)$, and transforming into an interaction picture where $H_{\text{field}}$ is rotated out, we obtain the interaction picture Hamiltonian \cite{Ko_2022, Combes_2017_review}
\begin{equation}
    H(t) = H_{\text{sys}} +\sum_l - ia_l(t)L_l^\dagger + ia_l^\dagger(t)L_l.
\label{Eq:H_int_state_perturbation}
\end{equation}
The time-dependent field operator $a_l(t)$ is defined as \cite{Loudon_2000_book}
\begin{equation}
    a_l(t) = \int \frac{d\omega}{\sqrt{2\pi}} \, a_l(\omega) e^{-i\omega t}, 
\end{equation}
and it satisfies the usual bosonic commutation relations: $[a_l(t), a_{l'}(t')]=0$ and $[a_l(t), a^\dagger_{l'}(t')]=\delta_{l,l'}\delta(t-t')$.
\par
We define the time evolution operator $U(t)$ as the solution to the Schrodinger equation
\begin{equation}
    dU(t)/dt = -iH(t)U(t)
\label{Eq:Schrodinger_eqn_modified_interaction_picture}
\end{equation}
with the initial condition $U(0)=1$. 
The time derivative of the density matrix of the system+field state $\rho_{\text{tot}}(t) = U(t)\rho_{\text{tot}}(0)U^\dagger(t)$ is given by 
\begin{equation}
    \frac{d}{dt}\rho_{\text{tot}}(t) = -i\big[H(t), \rho_{\text{tot}}(t)\big].
\label{Eq:von_Neumann_modified_interaction_picture}
\end{equation}
\par
An important identity regarding $U(t)$ is the commutation relation  
\begin{equation}
    [a_l(t), U(t)] = \frac{1}{2}L_lU(t),
\label{Eq:a_U_commutator}
\end{equation}
which is a consequence of the input-output relation. Derivations of Eq. (\ref{Eq:a_U_commutator}) can be found in Refs. \cite{Combes_2017_review}, \cite{Ko_2022}, and the Supplementary Material.

\section{Conventional perturbative expansion}
Before presenting the derivation of the normal-ordered expansion, we first briefly review the conventional perturbative expansion. We expand the reduced system state $\rho_{\text{sys}}(t)$ in orders of $a(t)$ and $a^\dagger(t)$, since the light-matter coupling strength is much weaker than the system Hamiltonian, i.e.,
\begin{equation}
    |H_{\text{sys}}|\gg|- ia(t)L^\dagger + ia^\dagger(t)L|.
\end{equation} 
The Liouville-von Neumann equation (Eq. (\ref{Eq:von_Neumann_modified_interaction_picture})) can be written in superoperator form as
\begin{equation}
    \frac{d}{dt}\rho_{\text{tot}}(t) = \mathcal{K}'\rho_{\text{tot}}(t) + \mathcal{L}(t)\rho_{\text{tot}}(t),
\label{Eq:von_Neumann_modified_interaction_picture_superoperator}
\end{equation}
where the superoperators $\mathcal{K}'$ and $\mathcal{L}$ are defined as
\begin{equation}
    \mathcal{K}' = -i[H_{\text{sys}}, \bullet]
\label{Eq:super_K_prime_definition}
\end{equation}
and
\begin{equation}
    \mathcal{L}(t) = \sum_l [-a_l(t)L_l^\dagger+a_l^\dagger(t)L_l,\bullet].
\label{Eq:super_L_definition}
\end{equation}
The superoperator notation with $\bullet$ means that the action of the superoperator on an operator is performed by substituting the operator into the $\bullet$. For example, $(-i[H_{\text{sys}}, \bullet])\rho = -i[H_{\text{sys}}, \rho]$. Solving Eq. (\ref{Eq:von_Neumann_modified_interaction_picture_superoperator}) formally, we have
\begin{equation}
    \rho_{\text{tot}}(t) = e^{\mathcal{K}'t}\rho_{\text{tot}}(0) + \int^t_0 dt_1 e^{\mathcal{K}'(t-t_1)}\mathcal{L}(t_1)\rho_{\text{tot}}(t_1).
\label{Eq:system_state_expansion_conventional_1}
\end{equation}
To clarify the notations, we note that $\mathcal{K}'(t-t')$ denotes the superoperator $\mathcal{K}'$ multiplied by $t-t'$, while $\mathcal{L}(t)$ means that the superoperator $\mathcal{L}$ is a function of time $t$.
Substituting Eq. (\ref{Eq:system_state_expansion_conventional_1}) into itself iteratively and then taking the partial trace $\text{Tr}_{\text{field}}$, we obtain the conventional expansion for $\rho_{\text{sys}}(t)$, i.e., 
\begin{align}
\begin{split}
    \rho_{\text{sys}}(t) = \text{Tr}_{\text{field}}\Big( &e^{\mathcal{K}'t} \rho_{\text{tot}}(0) \\
    &+\int^t_0 dt_1 \, e^{\mathcal{K}'(t-t_1)}\mathcal{L}(t_1) e^{\mathcal{K}'t_1}\rho_{\text{tot}}(0) \\
    & + \int^t_0 dt_2 \int^{t_2}_0 dt_1\, e^{\mathcal{K}'(t-t_2)}\mathcal{L}(t_2) e^{\mathcal{K}'(t_2-t_1)}\mathcal{L}(t_1) e^{\mathcal{K}'t_1}\rho_{\text{tot}}(0)\\
    & + \cdots \Big).
\label{Eq:conventional_expansion_system_state}
\end{split}
\end{align}
The perturbative expansion can be interpreted as periods of free evolution $e^{\mathcal{K}'(t-t')}$ interlaced with interactions $\mathcal{L}(t)$.
In this expansion, $\mathcal{K}'$ acts only on the system degrees of freedom. The field operators only appear in $\mathcal{L}$. The action of multiple $\mathcal{L}$ on $\rho_{\text{field}}(0)$ will result in expectation values of non-normal-ordered field operators. For example, the second order term in the expansion contains 8 types of field expectation values: $\langle a^\dagger(t_2)a^\dagger(t_1)\rangle$, $\langle a^\dagger(t_1)a^\dagger(t_2)\rangle$, $\langle a^\dagger(t_2)a(t_1)\rangle$, $\langle a^\dagger(t_1)a(t_2)\rangle$, $\langle a(t_2)a^\dagger(t_1)\rangle$, $\langle a(t_1)a^\dagger(t_2)\rangle$, $\langle a(t_2)a(t_1)\rangle$, and $\langle a(t_1)a(t_2)\rangle$. Two of these expectation values are not normal-ordered. 
\par
It is interesting to compare the perturbative expansions of $\rho_{\text{sys}}(t)$ under the semi-classical approximation of the light-matter interaction to the expansion of Eq. (\ref{Eq:conventional_expansion_system_state}) under a quantum mechanical coherent state input, which resembles classical coherent light.
Under the semi-classical approximation, one replaces the field operators $a(t)$ and $a^\dagger(t)$ in Eq. (\ref{Eq:conventional_expansion_system_state}) with the classical complex-valued amplitudes $\alpha(t)$ and $\alpha^*(t)$, which are proportional to the classical electric field amplitudes $E(t)$ and $E^*(t)$.
Under a quantum mechanical coherent state input, the field expectation values in Eq. (\ref{Eq:conventional_expansion_system_state}) is evaluated with respect to the coherent state. For the normal-ordered field expectation values, one can replace the field operators $a(t)$ and $a^\dagger(t)$ with the complex-valued coherent state amplitudes $\alpha(t)$ and $\alpha^*(t)$.
However, for the non-normal-ordered field expectation values, this replacement is not correct. To obtain the correct expectation values, one needs to account for the commutation relation of the field operators.
Therefore, the difference between $\rho_{\text{sys}}(t)$ under the semi-classical approximation and that under a quantum mechanical coherent state input lies in the non-normal-ordered field expectation values. 
From the conventional expansion of Eq. (\ref{Eq:conventional_expansion_system_state}), it is not easy to extract the quantum correction to the semi-classical treatment. We will see that the normal-ordered expansion of $\rho_{\text{sys}}(t)$ allows one to identify the quantum correction to the semi-classical treatment, even in the strong light-matter coupling regime.

\section{Normal-ordered perturbative expansion}
\label{Sec:normal_ordered_expansion}
Now, we derive the normal-ordered perturbative expansion of $\rho_{\text{sys}}(t)$. We will assume only one spatial mode to simplify the notations. Then we generalize the result to multiple spatial modes.

\par
Using the identity of Eq. (\ref{Eq:a_U_commutator}), we can write the Schrodinger equation (Eq. (\ref{Eq:Schrodinger_eqn_modified_interaction_picture})) in a normal-ordered form as
\begin{align}
\begin{split}
    \frac{dU(t)}{dt} &= -iH(t)U(t)\\
    &= -iH_{\text{sys}} U(t) - L^\dagger a(t)U(t) + a^\dagger(t)LU(t) \\
    &= -iH_{\text{sys}} U(t) - L^\dagger \big(\frac{1}{2}LU(t)+U(t)a(t)\big) + a^\dagger(t)LU(t)\\
    &= \big(-iH_{\text{sys}}-\frac{1}{2}L^\dagger L\big) U(t) - L^\dagger U(t)a(t) + a^\dagger(t)LU(t).
\label{Eq:normal_oredered_dU_dt}
\end{split}
\end{align}
This allows us to express the time derivative of the reduced time evolution superoperator as
\begin{align}
\begin{split}
    &\frac{d}{dt}\text{Tr}_{\text{field}}\Big( U(t)\bullet U^\dagger(t) \Big) \\
    & = \text{Tr}_{\text{field}}\bigg( -i\big[H_{\text{sys}}, U(t)\bullet U^\dagger(t)\big] - \frac{1}{2} \big\{L^\dagger L, U(t)\bullet U^\dagger(t)\big\} \\
    &\qquad\qquad -L^\dagger U(t)a(t)\bullet U^\dagger(t) - U(t) \bullet a^\dagger(t)U^\dagger(t)L \\
    &\qquad\qquad +U(t)\bullet U^\dagger(t)L^\dagger a(t) + a^\dagger(t)LU(t)\bullet U^\dagger(t) \bigg).
\label{Eq:dt_reduced_time_evol_super_1}
\end{split}
\end{align}
The notation $\{A,B\} = AB+BA$ denotes the anticommutator.
\par
We want to turn this expression into to a form such that all $a(t)$ appear on the left of $\bullet$ as $U(t)a(t)\bullet U^\dagger(t)$ and all $a^\dagger(t)$ appear on the right of $\bullet$ as $U(t)\bullet a^\dagger(t) U^\dagger(t)$. The reason for this arrangement will become clear when we later substitute the $\bullet$ with the initial state $\rho_{\text{tot}}(0)=\rho_{\text{sys}}(0)\otimes\rho_{\text{field}}(0)$, so that the field partial trace takes the form
\begin{equation}
    \text{Tr}_{\text{field}}\big(a(t'_1)\cdots a(t'_m)\rho_{\text{tot}}(0)a^\dagger(t''_n)\cdots a^\dagger(t''_1)\big).
\label{Eq:normal_ordered_form_Tr_field}
\end{equation}
Since the partial trace is invariant under cyclic permutations of field operators, Eq. (\ref{Eq:normal_ordered_form_Tr_field}) becomes
\begin{equation}
    \big\langle a^\dagger(t''_n)\cdots a^\dagger(t''_1)a(t'_1)\cdots a(t'_m)\big\rangle \rho_{\text{sys}}(0),
\label{Eq:normal_ordered_expectation}
\end{equation}
where the normal-ordered expectation value $\langle \cdots \rangle$ is evaluated with respect to $\rho_{\text{field}}(0)$.
Therefore, the field operator ordering of Eq. (\ref{Eq:normal_ordered_form_Tr_field}) ensures that only normal-ordered field expectation values will be evaluated.
\par
The final two terms on the right hand side of Eq. (\ref{Eq:dt_reduced_time_evol_super_1}) do not take the desired form and need to be converted into the normal-ordered form. Under the partial trace $\text{Tr}_{\text{field}}$, the field operators $a(t)$ and $a^\dagger(t)$ can be permuted cyclically. Therefore, the second to the last term on the right hand side of Eq. (\ref{Eq:dt_reduced_time_evol_super_1}) becomes
\begin{align}
\begin{split}
    \text{Tr}_{\text{field}}\big(U(t)\bullet U^\dagger(t)L^\dagger a(t)\big) &= \text{Tr}_{\text{field}}\big(a(t)U(t)\bullet U^\dagger(t)L^\dagger \big) \\
    &= \text{Tr}_{\text{field}}\big(U(t)a(t)\bullet U^\dagger(t)L^\dagger + \frac{1}{2} LU(t)\bullet U^\dagger(t)L^\dagger \big),
\label{Eq:dt_reduced_time_evol_super_simplifying_second_to_last}
\end{split}
\end{align}
where we have applied the commutation relation of Eq. (\ref{Eq:a_U_commutator}) to obtain the last equality. The last term on the right hand side of Eq. (\ref{Eq:dt_reduced_time_evol_super_1}) is simply the Hermitian conjugate of Eq. (\ref{Eq:dt_reduced_time_evol_super_simplifying_second_to_last}).
\par
Now, Eq. (\ref{Eq:dt_reduced_time_evol_super_1}) can be put in the desired form as
\begin{align}
\begin{split}
    &\frac{d}{dt}\text{Tr}_{\text{field}}\Big( U(t)\bullet U^\dagger(t) \Big) \\
    & = \text{Tr}_{\text{field}}\bigg( -i\big[H_{\text{sys}}, U(t)\bullet U^\dagger(t)\big] - \frac{1}{2} \big\{L^\dagger L, U(t)\bullet U^\dagger(t)\big\} + L U(t)\bullet U^\dagger(t) L^\dagger\\
    &\qquad\qquad -\big[L^\dagger,  U(t)a(t)\bullet U^\dagger(t)\big] + \big[L,U(t)\bullet a^\dagger(t) U^\dagger(t)\big]\bigg).
\label{Eq:dt_reduced_time_evol_super_2}
\end{split}
\end{align}
This should be compared with the conventional result 
\begin{align}
\begin{split}
    &\frac{d}{dt}\text{Tr}_{\text{field}}\Big( U(t)\bullet U^\dagger(t) \Big) \\
    & = \text{Tr}_{\text{field}}\bigg( -i\big[H_{\text{sys}}, U(t)\bullet U^\dagger(t)\big] \\
    &\qquad\qquad -\big[a(t)L^\dagger,  U(t)\bullet U^\dagger(t)\big] + \big[a^\dagger(t)L,U(t)\bullet  U^\dagger(t)\big]\bigg),
\label{Eq:dt_reduced_time_evol_super_3}
\end{split}
\end{align}
where no normal-ordering is applied. Comparing Eqs. (\ref{Eq:dt_reduced_time_evol_super_2}) and (\ref{Eq:dt_reduced_time_evol_super_3}), we see that by ordering $a(t)$ and $a^\dagger(t)$ into the forms $U(t)a(t)\bullet U^\dagger(t)$ and $U(t)\bullet a^\dagger(t) U^\dagger(t)$, a Lindblad dissipator term (i.e., $- \frac{1}{2} \{L^\dagger L, U(t)\bullet U^\dagger(t)\} + L U(t)\bullet U^\dagger(t) L^\dagger$) has to be added. 
\par
Since system operators can be taken out of the partial trace over field, Eq. (\ref{Eq:dt_reduced_time_evol_super_2}) can be re-written as
\begin{align}
\begin{split}
    &\frac{d}{dt}\text{Tr}_{\text{field}}\Big( U(t)\bullet U^\dagger(t) \Big) \\
    & = \mathcal{K}\Big( \text{Tr}_{\text{field}}\big(U(t)\bullet U^\dagger(t)\big)\Big) \\
    & \quad -\Big[L^\dagger,  \text{Tr}_{\text{field}}\big(U(t)a(t)\bullet U^\dagger(t)\big)\Big] + \Big[L,\text{Tr}_{\text{field}}\big(U(t)\bullet a^\dagger(t) U^\dagger(t)\big)\Big],
\label{Eq:dt_reduced_time_evol_super_4}
\end{split}
\end{align}
where $\mathcal{K}$ is the superoperator
\begin{equation}
    \mathcal{K}=-i[H_{\text{sys}}, \bullet] - \frac{1}{2}\{L^\dagger L \bullet\} + L\bullet L^\dagger.
\label{Eq:super_K_definition}
\end{equation}
Solving Eq. (\ref{Eq:dt_reduced_time_evol_super_4}) formally, we have
\begin{align}
\begin{split}
    \text{Tr}_{\text{field}}\Big(U(t)\bullet U^\dagger(t)\Big) =& e^{\mathcal{K}t} \text{Tr}_{\text{field}}(\bullet) + \int^t_0 dt_1 \,e^{\mathcal{K}(t-t_1)} \\
    &\bigg(\Big[-L^\dagger,  \text{Tr}_{\text{field}}\big(U(t_1)a(t_1)\bullet U^\dagger(t_1)\big)\Big] + \Big[L,\text{Tr}_{\text{field}}\big(U(t_1)\bullet a^\dagger(t_1) U^\dagger(t_1)\big)\Big]\bigg).
\label{Eq:solution_dt_reduced_time_evol}
\end{split}
\end{align}
Applying Eq. (\ref{Eq:solution_dt_reduced_time_evol}) to the combined initial state $\rho_{\text{tot}}(0)$, we have
\begin{align}
\begin{split}
    \rho_{\text{sys}}(t) &= \text{Tr}_{\text{field}}\Big(U(t)\rho_{\text{tot}}(0) U^\dagger(t)\Big) \\
    &=e^{\mathcal{K}t} \text{Tr}_{\text{field}}\big(\rho_{\text{tot}}(0)\big) + \int^t_0 dt_1 \,e^{\mathcal{K}(t-t_1)}\\
    &\quad \bigg(\Big[-L^\dagger,  \text{Tr}_{\text{field}}\big(U(t_1)a(t_1)\rho_{\text{tot}}(0) U^\dagger(t_1)\big)\Big] + \Big[L,\text{Tr}_{\text{field}}\big(U(t_1)\rho_{\text{tot}}(0) a^\dagger(t_1) U^\dagger(t_1)\big)\Big]\bigg).
\label{Eq:normal_ordered_expansion_derivation_1}
\end{split}
\end{align}
We can then expand $\text{Tr}_{\text{field}}\big(U(t_1)a(t_1)\rho_{\text{tot}}(0) U^\dagger(t_1)\big)$ and $\text{Tr}_{\text{field}}\big(U(t_1)\rho_{\text{tot}}(0)a^\dagger(t_1) U^\dagger(t_1)\big)$ in the integrand by applying Eq. (\ref{Eq:solution_dt_reduced_time_evol}) to $a(t_1)\rho_{\text{tot}}(0)$ and $\rho_{\text{tot}}(0)a^\dagger(t_1)$. Then $\rho_{\text{sys}}(t)$ becomes
\begin{align}
\begin{split}
    \rho_{\text{sys}}(t) = &\, e^{\mathcal{K}t}\text{Tr}_{\text{field}}\big(\rho_{\text{tot}}(0)\big) \\
    &+ \int^t_0 dt_1\, e^{\mathcal{K}(t-t_1)} \bigg( \Big[ - L^\dagger , e^{\mathcal{K}t_1}\text{Tr}_{\text{field}}\big(a(t_1)\rho_{\text{tot}}(0)\big)\Big] \\
    & \qquad\qquad\qquad\qquad + \Big[ L , e^{\mathcal{K}t_1}\text{Tr}_{\text{field}}\big(\rho_{\text{tot}}(0)a^\dagger(t_1)\big)\Big]\bigg)\\
    & + \int^t_0 dt_2 \int^{t_2}_0 dt_1\, e^{\mathcal{K}(t-t_2)} \bigg( \Big[-L^\dagger, e^{\mathcal{K}(t_2-t_1)}\Big[-L^\dagger, \text{Tr}_{\text{field}}\big(U(t_1)a(t_1)a(t_2)\rho_{\text{tot}}(0) U^\dagger(t_1)\big)\Big]\Big]\\
    & \qquad\qquad\qquad\qquad\qquad\quad + \Big[-L^\dagger, e^{\mathcal{K}(t_2-t_1)}\Big[L, \text{Tr}_{\text{field}}\big(U(t_1) a(t_2)\rho_{\text{tot}}(0)a^\dagger(t_1)) U^\dagger(t_1)\big)\Big]\Big] \\
    & \qquad\qquad\qquad\qquad\qquad\quad + \Big[L, e^{\mathcal{K}(t_2-t_1)}\Big[-L^\dagger, \text{Tr}_{\text{field}}\big(U(t_1) a(t_1)\rho_{\text{tot}}(0)a^\dagger(t_2) U^\dagger(t_1)\big)\Big]\Big] \\
    & \qquad\qquad\qquad\qquad\qquad\quad + \Big[L, e^{\mathcal{K}(t_2-t_1)}\Big[L, \text{Tr}_{\text{field}}\big( U(t_1)\rho_{\text{tot}}(0)a^\dagger(t_2)a^\dagger(t_1) U^\dagger(t_1)\big)\Big]\Big] \bigg).
\label{Eq:normal_ordered_expansion_system_state_2nd_iter}
\end{split}
\end{align}
In this expansion, we have new terms like $\text{Tr}_{\text{field}}\big(U(t_1)a(t_1)a(t_2)\rho_{\text{tot}}(0) U^\dagger(t_1)\big)$, where there are now two field operators in the partial trace. We can apply Eq. (\ref{Eq:solution_dt_reduced_time_evol}) again to these new terms. Repeating this iterative procedure, we obtain the normal-ordered expansion for the reduced system state:
\begin{align}
\begin{split}
    \rho_{\text{sys}}(t) = &\, e^{\mathcal{K}t}\text{Tr}_{\text{field}}\big(\rho_{\text{tot}}(0)\big) \\
    &+ \int^t_0 dt_1\, e^{\mathcal{K}(t-t_1)} \bigg( \Big[ - L^\dagger , e^{\mathcal{K}t_1}\text{Tr}_{\text{field}}\big(a(t_1)\rho_{\text{tot}}(0)\big)\Big] \\
    & \qquad\qquad\qquad\qquad + \Big[ L , e^{\mathcal{K}t_1}\text{Tr}_{\text{field}}\big(\rho_{\text{tot}}(0)a^\dagger(t_1)\big)\Big]\bigg)\\
    & + \int^t_0 dt_2 \int^{t_2}_0 dt_1\, e^{\mathcal{K}(t-t_2)} \bigg( \Big[-L^\dagger, e^{\mathcal{K}(t_2-t_1)}\Big[-L^\dagger, e^{\mathcal{K}t_1}\text{Tr}_{\text{field}}\big(a(t_1)a(t_2)\rho_{\text{tot}}(0) \big)\Big]\Big]\\
    & \qquad\qquad\qquad\qquad\qquad\quad + \Big[-L^\dagger, e^{\mathcal{K}(t_2-t_1)}\Big[L, e^{\mathcal{K}t_1} \text{Tr}_{\text{field}}\big( a(t_2)\rho_{\text{tot}}(0)a^\dagger(t_1)) \big)\Big]\Big] \\
    & \qquad\qquad\qquad\qquad\qquad\quad + \Big[L, e^{\mathcal{K}(t_2-t_1)}\Big[-L^\dagger, e^{\mathcal{K}t_1} \text{Tr}_{\text{field}}\big( a(t_1)\rho_{\text{tot}}(0)a^\dagger(t_2) \big)\Big]\Big] \\
    & \qquad\qquad\qquad\qquad\qquad\quad + \Big[L, e^{\mathcal{K}(t_2-t_1)}\Big[L, e^{\mathcal{K}t_1}\text{Tr}_{\text{field}}\big( \rho_{\text{tot}}(0)a^\dagger(t_2)a^\dagger(t_1) \big)\Big]\Big] \bigg) \\
    &+ \cdots.
\label{Eq:normal_ordered_expansion_system_state_pre}
\end{split}
\end{align}
Notice that Eq. (\ref{Eq:solution_dt_reduced_time_evol}) ensures that all field operators in this expansion take the normal ordered form of Eq. (\ref{Eq:normal_ordered_form_Tr_field}).
\par
Since the superoperator $\mathcal{K}$ and the operator $L$ act only on the system degrees of freedom, we can pull out the $\text{Tr}_{\text{field}}$ to the front and apply this partial trace to all terms on the right hand side of Eq. (\ref{Eq:normal_ordered_expansion_system_state_pre}). 
Furthermore, because normal-ordering of the field operators is preserved throughout the expansion, we do not need to keep track of the exact ordering of the field operators during the expansion. We just need to re-order the field operators according to Eq. (\ref{Eq:normal_ordered_form_Tr_field}) at the end. 
Combining the steps described above, we can now re-express Eq. (\ref{Eq:normal_ordered_expansion_system_state_pre}) compactly as
\begin{align}
\begin{split}
    \rho_{\text{sys}}(t) = \text{Tr}_{\text{field}}\hat{\mathcal{N}}\Bigg( &e^{\mathcal{K}t}\rho_{\text{tot}}(0) \\
    &+ \int^t_0 dt_1\, e^{\mathcal{K}(t-t_1)} \mathcal{L}(t_1) e^{\mathcal{K}t_1}\rho_{\text{tot}}(0) \\
    &+ \int^t_0 dt_2 \int^{t_2}_0 dt_1\, e^{\mathcal{K}(t-t_2)}\mathcal{L}(t_2)e^{\mathcal{K}(t_2-t_1)}\mathcal{L}(t_1)e^{\mathcal{K}t_1}\rho_{\text{tot}}(0)\\
    & + \cdots \Bigg).
\label{Eq:normal_ordered_expansion_system_state}
\end{split}
\end{align}
This is the normal-ordered perturbative expansion for $\rho_{\text{sys}}(t)$.
The notation $\text{Tr}_{\text{field}}\hat{\mathcal{N}}$ means to normal-order the field operators according to Eq. (\ref{Eq:normal_ordered_form_Tr_field}), and then evaluate the partial trace over the field degrees of freedom.
\par
The derivation of Eq. (\ref{Eq:normal_ordered_expansion_system_state}) generalize straightforwardly to the case of multiple spatial modes of light (indexed by $l$). The final result takes the same form as Eq. (\ref{Eq:normal_ordered_expansion_system_state}), but with the superoperators $\mathcal{K}$ and $\mathcal{L}$ modified to sum over the photon modes, i.e.,
\begin{equation}
    \mathcal{K}=-i[H_{\text{sys}}, \bullet] +\sum_l\Big(- \frac{1}{2}\{L_l^\dagger L_l \bullet\} + L_l\bullet L_l^\dagger\Big)
\label{Eq:super_K_definition_many_modes}
\end{equation}
and
\begin{equation}
    \mathcal{L}(t) = \sum_l [-a_l(t)L_l^\dagger+a_l^\dagger(t)L_l,\bullet].
\label{Eq:super_L_definition_many_modes}
\end{equation}
\par
The key difference between the normal-ordered expansion and the conventional expansion is in their free evolution generator $\mathcal{K}$ and $\mathcal{K}'$ (see Eqs. (\ref{Eq:super_K_prime_definition}) and (\ref{Eq:super_K_definition_many_modes})). In the normal-ordered expansion, the free evolution $\mathcal{K}'$ has an additional Lindbladian spontaneous emission term. This is what is needed to normal-order the field expectation values in the conventional expansion.

\par
An interesting application of the normal-ordered expansion is for an $m$-photon Fock state input
\begin{equation}
    |\xi_m\rangle = \frac{1}{\sqrt{m!}}\Big(\int dt\,\xi(t)a^\dagger(t) \Big)^m|\text{vac}\rangle,
\label{Eq:m_photon_Fock_state_def}
\end{equation}
where the temporal profile $\xi(t)$ is normalized as $\int dt\, |\xi(t)|^2 = 1$, so that $\langle \xi_m|\xi_m\rangle = 1$. 
The expectation value of a string of normal-ordered field operators is nonzero only when the number of creation operators $a^\dagger(t)$ is equal to the number of annihilation operators $a(t)$ and when the number of annihilation operators is less than or equal to $m$. Specifically, the $2k$-point correlation function is
\begin{equation}
    \langle \xi_m|a^\dagger(t_1)\cdots a^\dagger(t_k)a(t'_k)\cdots a(t'_1)|\xi_m\rangle = \frac{m!}{(m-k)!} \xi^*(t_1)\cdots\xi^*(t_k)\xi(t'_k)\cdots \xi(t'_1)
\end{equation}
for $k=1,2,\cdots, m$.
All expansion terms of order higher than $2m$ vanish exactly. Therefore, the normal-ordered perturbative expansion truncates exactly at order $2m$. In other words, the truncated normal-ordered perturbative series becomes exact, regardless of the light-matter coupling strength. This is in contrast with the conventional perturbative expansion, which continues to the infinite order, and that the truncated expansion is a good approximation only under the weak light-matter coupling regime.
\par
The master equation for $\rho_{\text{sys}}(t)$ can be obtained by taking the derivative of Eq. (\ref{Eq:normal_ordered_expansion_system_state}). However, the master equation usually does not close onto itself, meaning that the expression of $d\rho_{\text{sys}}(t)/dt$ involves another auxiliary density matrix $\rho_1(t)$, and the time derivative of $\rho_1(t)$ then involves yet another auxiliary density matrix $\rho_2(t)$, and so on.
For some special input states, the corresponding master equations can be expressed in closed forms \cite{Gardiner1985, Baragiola_2012}. 
In the Supplementary Material, we re-derive the master equations for a coherent state, m-photon Fock state, and Markovian Gaussian state inputs by directly taking the derivative of Eq. (\ref{Eq:normal_ordered_expansion_system_state}). The derivations are performed under a single spatial photon mode for notational simplicity. Generalization to many spatial modes can be done straightforwardly by summing over the spatial modes.
We note that these master equations are exact in all light-matter coupling regimes, since they are derived from the full perturbative expansion, and are not approximated by truncating the perturbative expansion to a finite (usually second) order.

\section{Numerical evaluation of the perturbative expansions}
\label{Sec:numerical_perturbative_expansion}
We now study the numerical accuracy of the conventional expansion (Eq. (\ref{Eq:conventional_expansion_system_state})) vs. the normal-ordered expansion (Eq. (\ref{Eq:normal_ordered_expansion_system_state})).
We study the excitation by a coherent state, an m-photon Fock state, and a cat state. For the coherent state and the m-photon Fock state inputs, we compute the exact dynamics using the corresponding master equations (see Supplementary Material).
\par
The perturbative expansion is dominated by the low order terms when the magnitude of the perturbation $\int dt\, a(t)L^\dagger$ is much less than $1$.
To estimate the order of magnitude, we notice that since $\int dt\, \langle a^\dagger(t)a(t)\rangle$ is equal to the average number of photons $m$ in the pulse, if the pulse duration is $\sigma$, then $\langle a^\dagger(t)a(t)\rangle$ has an order of magnitude of $m/\sigma$ during the pulse. Therefore, we assign an order of magnitude of $\sqrt{m/\sigma}$ to $a(t)$. The system operator $L$ has an order of magnitude of $\sqrt{\gamma}$, where $\gamma$ is the spontaneous emission rate into the spatial mode of the field. This is because the spontaneous emission rate is given by the expectation value $\langle L^\dagger L\rangle$. Combining these order of magnitude estimates, and using the fact that the integrand in $\int dt\, a(t)L^\dagger$ contributes significantly only within the pulse duration $\sigma$, we conclude that the perturbation parameter $\int dt\, a(t)L^\dagger$ has an order of magnitude of $\sqrt{m\sigma\gamma}$. 
\par

We consider a matter system consisting of two electronic levels (labeled as ground state $|g\rangle$ and excited state $|e\rangle$) interacting with a bath (e.g., nuclear degrees of freedom) that gives rise to Lindbladian pure dephasing in the electronic subsystem \cite{rebentrost2009environment}. Tracing out the bath degrees of freedom and focus on the two electronic states, we model the reduced electronic system dynamics as
\begin{equation}
    \frac{d}{dt}\rho_{\text{el}} = -i\big[H'_{\text{el}}, \rho_{\text{el}}\big] +\gamma_d \Big(- \frac{1}{2}\big\{P^\dagger P, \rho_{\text{el}}\big\} + P \rho_{\text{el}} P^\dagger\Big),
\end{equation}
where $\gamma_d$ is the dephasing rate, and $P = |e\rangle\langle e|$. 
We take $H=\omega|e\rangle\langle e|$. The dipole de-excitation operator $L$ is taken to be $\sqrt{\gamma}|g\rangle\langle e|$, where $\gamma$ is the coupling strength between the system and the field. $\gamma$ is also the spontaneous emission rate into the field. 
We consider an excitation pulse with a normalized Gaussian temporal profile
\begin{equation}
    \xi'(t) = \frac{1}{(\pi\sigma^2)^{1/4}}e^{-\frac{(t-t_0)^2}{2\sigma^2}} e^{-i\omega t},
\end{equation}
which is centered at time $t_0$ with a pulse duration of $\sigma$. The carrier frequency $\omega$ is resonant with the energy splitting of the two electronic levels. $\xi'$ is normalized such that $\int dt\, |\xi'(t)|^2 = 1$. We set $t_0=5$ ps, $\sigma=1$ ps, and $\gamma_d=1 (\text{ps})^{-1}$.
\par
In the numerical calculation, we work in a rotating frame where the carrier frequency $\omega$ is rotated out. In this frame, the temporal profile becomes 
\begin{equation}
    \xi(t) = \frac{1}{(\pi\sigma^2)^{1/4}}e^{-\frac{(t-t_0)^2}{2\sigma^2}},
\label{Eq:normalized_temporal_profile}
\end{equation}
and the Hamiltonian becomes $H_{\text{el}} = H'_{\text{el}}-\omega|e\rangle\langle e| = 0$.
 
\par
The initial state is set to be in the ground state $|g\rangle\langle g|$. We will compute the excited state population $\langle e|\rho_{\text{sys}}(t)|e\rangle$ as a function of time. The change from the initial ground state $|g\rangle\langle g|$ to the excited state population $|e\rangle\langle e|$ requires at least two interactions (i.e., $L^\dagger |g\rangle\langle g|L = |e\rangle\langle e|$). Therefore, we need to perform the perturbative expansion to at least the second order to see nonzero excited state population. 
\par
In the conventional expansion (Eq. (\ref{Eq:conventional_expansion_system_state})), the superoperator that generates free evolution in the electronic degrees of freedom is
\begin{equation}
    \mathcal{K}' = -i\big[H_{\text{el}},\bullet\big] +\gamma_d \Big(- \frac{1}{2}\{P^\dagger P, \bullet\} + P \bullet  P^\dagger\Big).
\end{equation}
In the normal-ordered expansion (Eq. (\ref{Eq:normal_ordered_expansion_system_state})), the superoperator that generates free evolution is
\begin{align}
\begin{split}
    \mathcal{K} = -i\big[H_{\text{el}},\bullet\big] &+\gamma_d \Big(- \frac{1}{2}\{P^\dagger P, \bullet\} + P \bullet  P^\dagger\Big)\\
    &- \frac{1}{2}\{L^\dagger L, \bullet\} + L \bullet  L^\dagger.
\end{split}
\end{align}
\par
We perform perturbative expansions up to the 6-th order. For the conventional expansion, we use Wick's theorem (or the commutation relation) to express the non-normal-ordered field correlation functions as sums of normal-ordered correlations, resulting in many additional terms. Explicit expressions for the expansion terms are given in the Supplementary Material.

\subsection{Coherent state input}
\begin{figure}[h]
    \centering
    \includegraphics[scale=0.8]{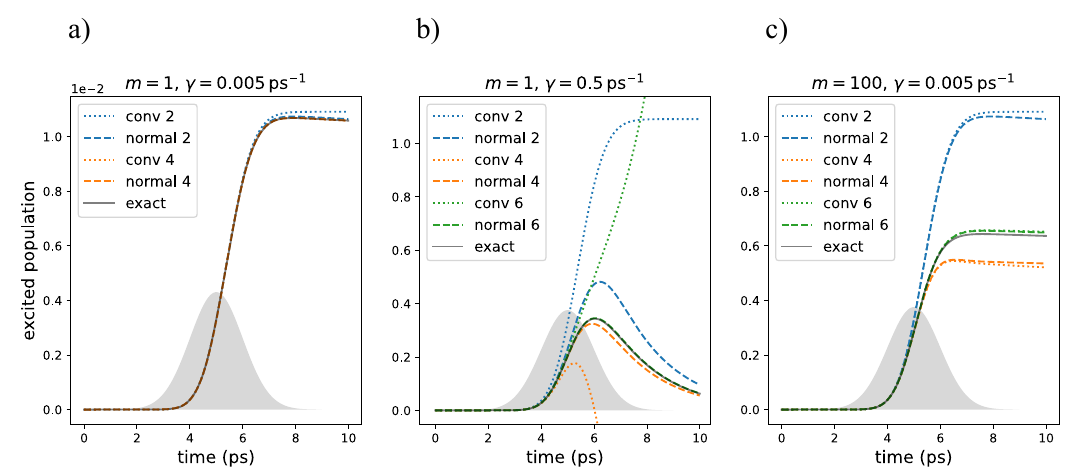}
    \caption{Comparing the conventional perturbative expansion with the normal-ordered perturbative expansion under coherent state input light in three different parameter regimes. In all parameter regimes, the normal-ordered expansions are more accurate than the conventional expansions. In the legends, ``conv $n$" means $n$-th order conventional expansion. ``normal $n$" means $n$-th order normal-ordered expansion.}
    \label{fig:coherent_expansion}
\end{figure}
A coherent state with coherent amplitude $\alpha(t)$ is defined as
\begin{equation}
    |\alpha(t)\rangle = \exp\Big( \int dt\, \alpha(t)a^\dagger(t) - \alpha^*(t)a(t)\Big)|\text{vac}\rangle.
\label{Eq:coherent_state_definition}
\end{equation}
We take the the coherent state amplitude $\alpha(t)=\sqrt{m}\xi(t)$, where $m$ is the average photon number, and $\xi(t)$ is given by Eq. (\ref{Eq:normalized_temporal_profile}). 
In Fig. (\ref{fig:coherent_expansion}a), a coherent state input with an average photon number of 1 (i.e., $m=1$) is used. The perturbative strength $\sqrt{m\sigma\gamma}\approx 0.07$ is much less than 1, so both expansion methods show good agreements with the exact dynamics, with the normal-ordered expansion being closer to the exact dynamics. Both fourth order expansions show no visible difference to the exact dynamics. In Fig. (\ref{fig:coherent_expansion}b), the light-matter coupling strength $\gamma$ is now 100 times larger than in Fig. (\ref{fig:coherent_expansion}a), so that the perturbative strength $\sqrt{m\sigma\gamma}\approx 0.7$ is on the order of 1. The conventional expansion is not expected to work well under such perturbative strength. Indeed, we see that conventional expansions tend to diverge at large $t$, producing unphysical results where the excited probability becomes $>1$ or $<0$. On the other hand, the normal-ordered expansions remain close to the exact dynamics, and the 6-th order normal-ordered expansion almost matches the exact dynamics. In Fig. (\ref{fig:coherent_expansion}c), the number of photons is now 100, and the light-matter coupling strength $\gamma$ is 100 times weaker than in Fig. (\ref{fig:coherent_expansion}b), resulting in the same perturbative strength $\sqrt{m\sigma\gamma}$ as in Fig. (\ref{fig:coherent_expansion}b). Within the time range of $T=10$ ps, the spontaneous emission effect is on the order of $\gamma T=0.05 \ll 1$. Therefore, within this time range, the normal-ordered expansion is only slightly more accurate than the conventional expansion. At longer times, the normal-ordered expansion become significantly more accurate than the conventional expansion, since it accounts for the spontaneous emission effects properly.

\subsection{$m$-photon Fock state input}
\begin{figure}[h]
    \centering
    \includegraphics[scale=0.8]{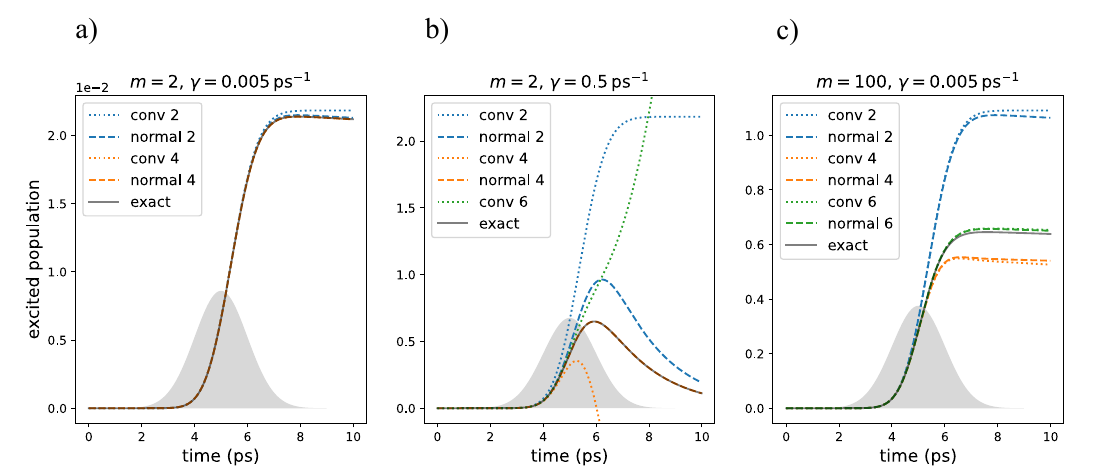}
    \caption{Comparing the conventional perturbative expansion with the normal-ordered perturbative expansion under $m$-photon Fock state input light in three different parameter regimes. The $2m$-th order normal-ordered expansion produces the exact result in all regimes. In all parameter regimes, the normal-ordered expansions are more accurate than the conventional expansions. In the legends, ``conv $n$" means $n$-th order conventional expansion. ``normal $n$" means $n$-th order normal-ordered expansion.}
    \label{fig:Fock_expansion}
\end{figure}
We define an $m$-photon Fock state input according to Eq. (\ref{Eq:m_photon_Fock_state_def}). 
Fig. (\ref{fig:Fock_expansion}) compares the two expansion methods to the exact Fock state master equation (see Supplementary Material). As discussed at the end of Sec. \ref{Sec:normal_ordered_expansion}, the normal-ordered expansion for an $m$-photon Fock state input truncates exactly at the $2m$-th order. In Figs. (\ref{fig:Fock_expansion}a) and (\ref{fig:Fock_expansion}b), the input light is a 2-photon Fock state, so the 4-th order normal-ordered expansion yields the exact result. In Fig. (\ref{fig:Fock_expansion}a), the perturbative parameter $\sqrt{m\sigma\gamma}=0.1$ is much less than 1. Similar to Fig. (\ref{fig:coherent_expansion}a), both perturbative expansion methods match the exact dynamics well, and the second order normal-ordered expansion is more accurate than the second order conventional expansion. In Fig. (\ref{fig:Fock_expansion}b), the light-matter coupling $\gamma$ is increase by 100 times from Fig. (\ref{fig:Fock_expansion}b). The perturbation parameter $\sqrt{m\sigma\gamma}=1$ is now on the order of 1, and the conventional perturbative expansions fail. On the other hand, the normal-ordered expansion produces the exact dynamics at the 4-th order. In Fig. (\ref{fig:Fock_expansion}), the photon number is increased to 100, and $\gamma$ is decreased by 100 times from Fig. (\ref{fig:Fock_expansion}c). Similar to Fig. (\ref{fig:coherent_expansion}c), the normal-ordered expansion only results in slight improvement from the conventional expansion. At longer times, the normal-ordered expansion becomes significantly more accurate than the conventional expansion. Since $m=100$, the exact Fock state master equations are differential equations that couple $101\times 101=10,201$ auxiliary density matrices. Therefore, for large $m$, perturbative expansions are the computationally cheaper alternatives for computing the reduced system state. 

\subsection{Cat state input}
Finally, we consider a cat state input, taken to be the superposition of two coherent states. 
Specifically, we shall consider two kinds of cat state, an even cat state $|\text{cat}_+\rangle$ and an odd cat state $|\text{cat}_-\rangle$, defined as
\begin{equation}
    |\text{cat}_+\rangle = C_+ (|\xi(t)\rangle + |-\xi(t)\rangle)
\end{equation}
and
\begin{equation}
    |\text{cat}_-\rangle = C_- (|\xi(t)\rangle - |-\xi(t)\rangle),
\end{equation}
where $|\xi(t)\rangle$ and $|-\xi(t)\rangle$ are single-photon coherent states with coherent amplitudes $\xi(t)$ and $-\xi(t)$. The overlap $\langle \xi(t)|-\xi(t)\rangle$ is $e^{-2}$, and the normalization constants are $C_\pm = 1/\sqrt{2\pm2e^{-2}}$. The normal-ordered correlation functions that will be used to evaluate the perturbative expansion terms are the following:
\begin{subequations}
\begin{equation}
    \langle \text{cat}_\pm| a^\dagger(t_2)a(t_1)|\text{cat}_\pm\rangle = \frac{1\mp e^{-2}}{1\pm e^{-2}} \xi^*(t_2)\xi(t_1)
\label{Eq:cat_correlation_2_point}
\end{equation}
\begin{equation}
    \langle \text{cat}_\pm| a^\dagger(t_4)a^\dagger(t_3)a(t_2)a(t_1)|\text{cat}_\pm\rangle = \xi^*(t_4)\xi^*(t_3)\xi(t_2)\xi(t_1)
\end{equation}
\begin{equation}
    \langle \text{cat}_\pm| a^\dagger(t_6)a^\dagger(t_5)a^\dagger(t_4)a(t_3)a(t_2)a(t_1)|\text{cat}_\pm\rangle = \frac{1\mp e^{-2}}{1\pm e^{-2}} \xi^*(t_6)\xi^*(t_5)\xi^*(t_4)\xi(t_3)\xi(t_2)\xi(t_1).
\end{equation}
\end{subequations}
\par
\begin{figure}
    \centering
    \includegraphics[scale=0.6]{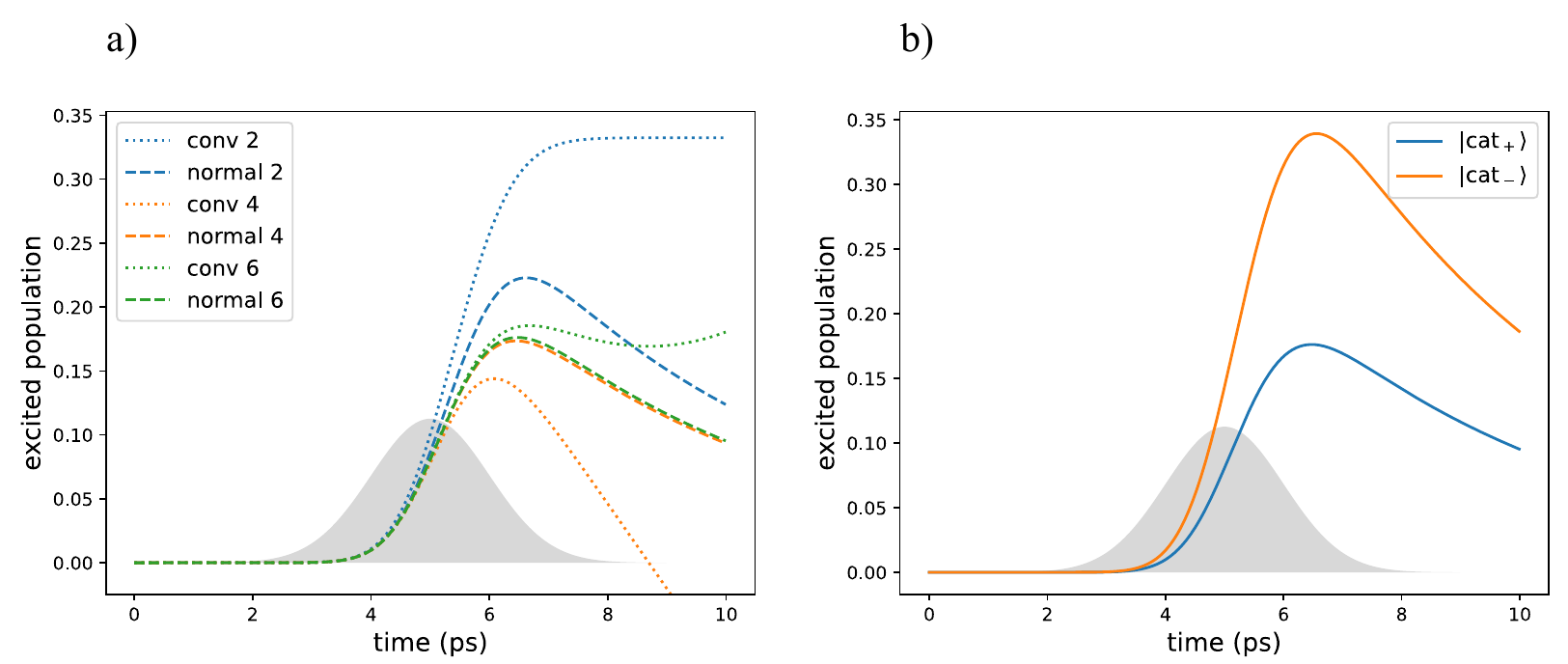}
    \caption{(a) Excitation probability under the even cat state $|\text{cat}_+\rangle$, computed with both expansion methods at different orders. The light-matter coupling strength $\gamma$ is taken to be $0.2 \text{ ps}^{-1}$. The normal-ordered expansion converges much more quickly than the conventional expansion does. }
    \label{fig:cat_calculation}
\end{figure}

Fig. (\ref{fig:cat_calculation}a) shows the system dynamics under the even cat state $|\text{cat}_+\rangle$ excitation, computed at different expansion orders. The conventional expansion is far from convergence even at the sixth order, while the normal-ordered expansion is already close to convergence at the fourth order. 
If one treats the cat state input semi-classically, one may conclude incorrectly that the total electric field is zero due to the destructive interference between the two coherent states of opposite phases, and therefore the excitation probability will remain $0$ for all time. 
In Fig. (\ref{fig:cat_calculation}b), we compare the excitation probability under $|\text{cat}_+\rangle$ and $|\text{cat}_-\rangle$. Under weak to intermediate light-matter coupling, $|\text{cat}_-\rangle$ results in higher excitation probability than $|\text{cat}_+\rangle$ does. This is because the second order expansion term dominates at weak to intermediate light-matter coupling, and the two-point correlation function (see Eq. (\ref{Eq:cat_correlation_2_point})) of $|\text{cat}_-\rangle$ is $(1+e^{-2})^2/(1-e^{-2})^2\approx 1.7$ times larger than that of $|\text{cat}_+\rangle$.

\section{Conclusion}
We have shown that the normal-ordered perturbative expansion for the reduced system state can be significantly more accurate and simpler to use than the conventional perturbative expansion. The normal-ordered expansion is useful in studying the reduced matter system dynamics under various input photon states when the corresponding master equations are not known. Furthermore, we have shown that the normal-ordered expansion can serve as a unified starting point to derive various master equations.

\section*{Acknowledgements}
L.K. was supported by the Kavli Energy NanoScience Institute (ENSI) Philomathia graduate fellowship. This project was supported by the Photosynthetic Systems program of U.S. Department of Energy, Office of Science, Basic Energy Sciences, within the Division of Chemical Sciences, Geosciences, and Biosciences, under Award No. DESC0019728.

\newpage
\bibliographystyle{unsrt}
\typeout{}
\bibliography{references.bib}

@article{gough2006quantum,
  title={Quantum white noise and the master equation for Gaussian reference states},
  author={Gough, John},
  journal={arXiv preprint quant-ph/0609040},
  year={2006}
}

@article{mitra2022quantum,
  title={Quantum control of molecules for fundamental physics},
  author={Mitra, D and Leung, KH and Zelevinsky, T},
  journal={Phys. Rev. A},
  volume={105},
  number={4},
  pages={040101},
  year={2022},
  publisher={APS}
}

@article{qi2021manipulation,
  title={Manipulation of matter with shaped-pulse light field and its applications},
  author={Qi, Hongxia and Lian, Zhenzhong and Fei, Dehou and Chen, Zhou and Hu, Zhan},
  journal={Adv. Phys.: X},
  volume={6},
  number={1},
  pages={1949390},
  year={2021},
  publisher={Taylor \& Francis}
}

@article{gu2022wave,
  title={Wave packet control and simulation protocol for entangled two-photon absorption of molecules},
  author={Gu, Bing and Keefer, Daniel and Mukamel, Shaul},
  journal={J. Chem. Theory Comput.},
  volume={18},
  number={1},
  pages={406--414},
  year={2022},
  publisher={ACS Publications}
}

@article{beyvers2006optimal,
  title={Optimal control in a dissipative system: Vibrational excitation of {CO}/ {C}u (100) by {IR} pulses},
  author={Beyvers, Stephanie and Ohtsuki, Yukiyoshi and Saalfrank, Peter},
  journal={J. Chem. Phys.},
  volume={124},
  number={23},
  year={2006},
  publisher={AIP Publishing}
}

@article{brumer1986control,
  title={Control of unimolecular reactions using coherent light},
  author={Brumer, Paul and Shapiro, Moshe},
  journal={Chem. Phys. Lett.},
  volume={126},
  number={6},
  pages={541--546},
  year={1986},
  publisher={Elsevier}
}

@article{schlawin2017theory,
  title={Theory of coherent control with quantum light},
  author={Schlawin, Frank and Buchleitner, Andreas},
  journal={New J. Phys.},
  volume={19},
  number={1},
  pages={013009},
  year={2017},
  publisher={IOP Publishing}
}

@article{warner2025coherent,
  title={Coherent control of a superconducting qubit using light},
  author={Warner, Hana K and Holzgrafe, Jeffrey and Yankelevich, Beatriz and Barton, David and Poletto, Stefano and Xin, CJ and Sinclair, Neil and Zhu, Di and Sete, Eyob and Langley, Brandon and others},
  journal={Nat. Phys.},
  volume={21},
  number={5},
  pages={831--838},
  year={2025},
  publisher={Nature Publishing Group UK London}
}

@article{makri1998quantum,
  title={Quantum dissipative dynamics: A numerically exact methodology},
  author={Makri, Nancy},
  journal={J. Phys. Chem. A},
  volume={102},
  number={24},
  pages={4414--4427},
  year={1998},
  publisher={ACS Publications}
}

@article{gross2022master,
  title={Master equations and quantum trajectories for squeezed wave packets},
  author={Gross, Jonathan A and Baragiola, Ben Q and Stace, TM and Combes, Joshua},
  journal={Phys. Rev. A},
  volume={105},
  number={2},
  pages={023721},
  year={2022},
  publisher={APS}
}

@article{amgar2019higher,
  title={Higher-order photon correlation as a tool to study exciton dynamics in quasi-2{D} nanoplatelets},
  author={Amgar, Daniel and Yang, Gaoling and Tenne, Ron and Oron, Dan},
  journal={Nano Lett.},
  volume={19},
  number={12},
  pages={8741--8748},
  year={2019},
  publisher={ACS Publications}
}

@article{kimble2008quantum,
  title={The quantum internet},
  author={Kimble, H Jeff},
  journal={Nature},
  volume={453},
  number={7198},
  pages={1023--1030},
  year={2008},
  publisher={Nature Publishing Group}
}

@article{Ko_2022,
author = {Ko,L.  and Cook,R. L.  and Whaley,K. B. },
title = {Dynamics of photosynthetic light harvesting systems interacting with N-photon Fock states},
journal = {J. Chem. Phys.},
volume = {156},
number = {24},
pages = {244108},
year = {2022},
doi = {10.1063/5.0082822},

URL = { 
        https://doi.org/10.1063/5.0082822
    
},
eprint = { 
        https://doi.org/10.1063/5.0082822
    
}

}

@book{Loudon_2000_book,
  title={The quantum theory of light},
  author={Loudon, R.},
  year={2000},
  publisher={Oxford University Press}
}

@article{Mukamel_Rev_Mod_Phys,
  title = {Nonlinear optical signals and spectroscopy with quantum light},
  author = {Dorfman, K. E. and Schlawin, F. and Mukamel, S.},
  journal = {Rev. Mod. Phys.},
  volume = {88},
  number = {4},
  pages = {045008},
  numpages = {67},
  year = {2016},
  publisher = {American Physical Society},
  doi = {10.1103/RevModPhys.88.045008},
  url = {https://link.aps.org/doi/10.1103/RevModPhys.88.045008}
}

@article{Schlawin_2017_tutorial,
doi = {10.1088/1361-6455/aa8a7a},
url = {https://dx.doi.org/10.1088/1361-6455/aa8a7a},
year = {2017},
publisher = {IOP Publishing},
volume = {50},
number = {20},
pages = {203001},
author = {Schlawin, F.},
title = {Entangled photon spectroscopy},
journal = {J. Phys. B: At., Mol. Opt. Phys.}
}

@article{Gardiner1985,
  title={Input and output in damped quantum systems: Quantum stochastic differential equations and the master equation},
  author={Gardiner, C. W. and Collett, M. J.},
  journal={Phys. Rev. A},
  volume={31},
  number={6},
  pages={3761},
  year={1985},
  publisher={APS}
}

@article{Baragiola_2012,
  title = {{N}-photon wave packets interacting with an arbitrary quantum system},
  author = {Baragiola, B. Q. and Cook, R. L. and Bra\ifmmode \acute{n}\else \'{n}\fi{}czyk, A. M. and Combes, J.},
  journal = {Phys. Rev. A},
  volume = {86},
  number = {1},
  pages = {013811},
  numpages = {18},
  year = {2012},
  publisher = {American Physical Society},
  doi = {10.1103/PhysRevA.86.013811},
  url = {https://link.aps.org/doi/10.1103/PhysRevA.86.013811}
}

@book{gardiner_zoller_quantum_noise,
  title={Quantum noise: a handbook of Markovian and non-Markovian quantum stochastic methods with applications to quantum optics},
  author={Gardiner, C. and Zoller, P.},
  year={2004},
  publisher={Springer Science \& Business Media}
}

@book{Mukamel_book,
  title={Principles of Nonlinear Optical Spectroscopy},
  author={Mukamel, S.},
  isbn={9780195092783},
  lccn={gb95087868},
  series={Oxford series in optical and imaging sciences},
  url={https://books.google.com/books?id=k_7uAAAAMAAJ},
  year={1995},
  publisher={Oxford University Press}
}

@article{lupton2021review,
  title={Photon correlations probe the quantized nature of light emission from optoelectronic materials},
  author={Lupton, J. M. and Vogelsang, J.},
  journal={Appl. Phys. Rev.},
  volume={8},
  number={4},
  year={2021},
  pages = {041302},
  publisher={AIP Publishing}
}

@article{Combes_2017_review,
author = {Combes, J. and Kerckhoff, J. and Sarovar, M.},
title = {The {SLH} framework for modeling quantum input-output networks},
journal = {Adv. Phys.: X},
volume = {2},
number = {3},
pages = {784-888},
year  = {2017},
publisher = {Taylor & Francis},
doi = {10.1080/23746149.2017.1343097},

URL = { 
        https://doi.org/10.1080/23746149.2017.1343097
},
eprint = { 
        https://doi.org/10.1080/23746149.2017.1343097
}
}

@book{wiseman_milburn_2009_book, place={Cambridge}, title={Quantum Measurement and Control}, DOI={10.1017/CBO9780511813948}, publisher={Cambridge University Press}, author={Wiseman, H. M. and Milburn, G. J.}, year={2009}}

@article{Tanimura_2020,
author = {Tanimura,Y. },
title = {Numerically exact approach to open quantum dynamics: The hierarchical equations of motion {(HEOM)}},
journal = {J. Chem. Phys.},
volume = {153},
number = {2},
pages = {020901},
year = {2020},
doi = {10.1063/5.0011599},
URL = { 
        https://doi.org/10.1063/5.0011599
},
eprint = { 
        https://doi.org/10.1063/5.0011599
}
}

@book{Rammer_book, place={Cambridge}, title={Quantum Field Theory of Non-equilibrium States}, DOI={10.1017/CBO9780511618956}, publisher={Cambridge University Press}, author={Rammer, J.}, year={2007}}

@article{li2023single,
  title={Single-photon absorption and emission from a natural photosynthetic complex},
  author={Li, Q. and Orcutt, K. and Cook, R. L. and Sabines-Chesterking, J. and Tong, A. L. and Schlau-Cohen, G. S. and Zhang, X. and Fleming, G. R. and Whaley, K. B.},
  journal={Nature},
  volume={619},
  number={7969},
  pages={300--304},
  year={2023},
  publisher={Nature Publishing Group UK London}
}

@article{rebentrost2009environment,
  title={Environment-assisted quantum transport},
  author={Rebentrost, P. and Mohseni, M. and Kassal, I. and Lloyd, S. and Aspuru-Guzik, A.},
  journal={New J. Phys.},
  volume={11},
  number={3},
  pages={033003},
  year={2009},
  publisher={IOP Publishing}
}

\newpage
\begin{center}
    \huge\bfseries Supplementary Material 
\end{center}
\vspace{1em} 

\setcounter{equation}{0}
\renewcommand{\theequation}{S.\arabic{equation}}
\setcounter{section}{0}                         
\renewcommand{\thesection}{S\arabic{section}}
\setcounter{figure}{0}                         
\renewcommand{\thefigure}{S\arabic{figure}} 

\section{Deriving the identity $[a_l(t), U(t)]=\frac{1}{2}L_l$}

We define $a(s,t)$ as $U^\dagger(t)a(s)U(t)$. Next, we take the partial derivative $\partial/\partial t$ using Eqs. (\ref{Eq:Schrodinger_eqn_modified_interaction_picture}) and (\ref{Eq:H_int_state_perturbation}) in the main text: (copied below)
\begin{equation}
    dU(t)/dt = -iH(t)U(t)
\end{equation}
\begin{equation}
    H(t) = H_{\text{sys}} +\sum_l - ia_l(t)L_l^\dagger + ia_l^\dagger(t)L_l.
\end{equation}
The result is
\begin{equation}
    \frac{\partial a(s,t)}{\partial t} = \delta(s-t) U^\dagger(t) L U(t),
\label{Eq:d_a_l_st_dt_modified_int}
\end{equation}
We let $s>0$, meaning that at time $t=0$, the plane of the photon field $a(s)$ has not interacted with the molecule located at the origin. Solving Eq. (\ref{Eq:d_a_l_st_dt_modified_int}) with the initial condition $a(s,0)=a(s)$, we obtain the input-output relation as
\begin{equation}
    a(s,t) = 
    \begin{cases}
        a(s)\qquad\qquad\qquad\qquad\,, t<s \\
        a(s) + \frac{1}{2}U^\dagger(t)LU(t) \quad\,\,, t=s \\
        a(s) + U^\dagger(t)L U(t) \qquad, t>s.
    \end{cases}
\end{equation}
When $t<s$, the Heisenberg-evolved field operator $a(s,t)$ is identified as the input field. When $t>s$, $a(s,t)$ is identified as the output field. When $t=s$, the factor of $1/2$ originates from cutting the delta function in half. In the case of $t=s$, we have 
\begin{equation}
    U^\dagger(t)a(t)U(t) = a(t) + \frac{1}{2} U^\dagger(t)LU(t).
\end{equation}
Left multiplying both sides of this equation by $U(t)$, we arrive at the commutation relation:
\begin{equation}
    [a_l(t), U(t)] = \frac{1}{2}L_l.
\end{equation}

\section{Deriving the coherent state master equation from the normal-ordered expansion}
\label{Sec:coherent_state_input_normal_ordering}
A coherent state with coherent amplitude $\alpha(t)$ is defined as
\begin{equation}
    |\alpha(t)\rangle = \exp\Big( \int dt\, \alpha(t)a^\dagger(t) - \alpha^*(t)a(t)\Big)|\text{vac}\rangle.
\label{Eq:coherent_state_definition_SM}
\end{equation}
It is an eigenstate of $a(t)$, i.e., $a(t)|\psi\rangle = \alpha(t)|\psi\rangle$. The expectation value of a normal-ordered string of $a(t)$ and $a^\dagger(t)$ (e.g., $\langle\psi|a^\dagger(t_4)a^\dagger(t_3)a(t_2)a(t_1)|\psi\rangle$) is obtained by replacing $a(t)$ and $a^\dagger(t)$ with $\alpha(t)$ and $\alpha^*(t)$, respectively. The example above would evaluate to $\alpha^*(t_4)\alpha^*(t_3)\alpha(t_2)\alpha(t_1)$.
\par
Taking the initial field state as the coherent state $|\alpha(t)\rangle$, we can simply replace the field operators $a(t)$ and $a^\dagger(t)$ in the normal-ordered perturbative expansion with the classical amplitudes $\alpha(t)$ and $\alpha^*(t)$. The normal-ordered expansion now becomes
\begin{align}
\begin{split}
    \rho_{\text{sys}}(t) =  &e^{\mathcal{K}t}\rho_{\text{sys}}(0) + \int^t_0 dt_1\, e^{\mathcal{K}(t-t_1)} \Tilde{\mathcal{L}}(t_1) e^{\mathcal{K}t_1}\rho_{\text{sys}}(0) \\
    &+ \int^t_0 dt_2 \int^{t_2}_0 dt_1\, e^{\mathcal{K}(t-t_2)}\Tilde{\mathcal{L}}(t_2)e^{\mathcal{K}(t_2-t_1)}\Tilde{\mathcal{L}}e^{\mathcal{K}t_1}\rho_{\text{sys}}(0)\\
    & + \cdots ,
\label{Eq:normal_ordered_expansion_system_state_coherent}
\end{split}
\end{align}
where $\Tilde{\mathcal{L}}(t) = [-\alpha(t)L^\dagger+\alpha^*(t)L,\bullet]$. Taking the time derivative of Eq. (\ref{Eq:normal_ordered_expansion_system_state_coherent}), we obtain the coherent state master equation \cite{Ko_2022, wiseman_milburn_2009_book}
\begin{equation}
    \frac{d}{dt}\rho_{\text{sys}}(t) = \mathcal{K}\rho_{\text{sys}}(t) + \Tilde{\mathcal{L}}(t)\rho_{\text{sys}}(t),
\end{equation}
or written more explicitly,
\begin{align}
\begin{split}
    \frac{d}{dt}\rho_{\text{sys}}(t) = &[-iH_{\text{sys}} - \alpha(t)L^\dagger + \alpha^*(t)L, \rho_{\text{sys}}(t)] \\
    &-\frac{1}{2}L^\dagger L \rho_{\text{sys}}(t) -\frac{1}{2}\rho_{\text{sys}}(t)L^\dagger L + L\rho_{\text{sys}}(t)L^\dagger.
\label{Eq:coherent_state_master}
\end{split}
\end{align}
The difference to the semi-classical master equation is in the Lindblad dissipator that describes spontaneous emission.

\section{Deriving the Fock state master equation from the normal-ordered expansion}

\label{sec:Fock_state_normal_ordered_expansion}
An m-photon Fock state $|\xi_m\rangle$ is defined as
\begin{equation}
    |\xi_m\rangle = \frac{1}{\sqrt{m!}}\Big(\int dt\,\xi(t)a^\dagger(t) \Big)^m|\text{vac}\rangle,
\label{Eq:m_photon_Fock_state_def_SM}
\end{equation}
where the temporal profile $\xi(t)$ is normalized as $\int dt\, |\xi(t)|^2 = 1$, so that $\langle \xi_m|\xi_m\rangle = 1$. 
\par

In the Fock state master equation \cite{Baragiola_2012, Ko_2022}, one defines auxiliary density matrices $\rho_{\mu, \nu}$ as
\begin{equation}
    \rho_{\mu,\nu} (t) = \text{Tr}_{\text{field}}\Big( U'(t) \big(\rho_{\text{sys}}(0)\otimes|\xi_\mu\rangle\langle \xi_\nu| \big)U'^\dagger(t)\Big).
\label{Eq:def_rho_mu_nu}
\end{equation}
In particular, $\rho_{m,m}(t)$ is the reduced system state under an $m$-photon Fock state input.
Note that, by definition, the initial condition for the auxiliary density matrices is $\rho_{\mu,\nu}(0) = \delta_{\mu,\nu}\rho_{\text{sys}}(0)$. 
Substituting the initial state $\rho_{\text{tot}}(0)=\rho_{\text{sys}}(0)\otimes|\xi_\mu\rangle\langle \xi_\nu| $ into the normal-ordered expansion, we have
\begin{align}
\begin{split}
    \rho_{\mu,\nu}(t) = \text{Tr}_{\text{field}}\hat{\mathcal{N}}\Big(R_{\mu,\nu}(t)\Big),
\label{Eq:rho_mu_nu_expansion}
\end{split}
\end{align}
where 
\begin{align}
\begin{split}
     R_{\mu,\nu}(t) =&\,e^{\mathcal{K}t}\rho_{\text{sys}}(0)\otimes|\xi_\mu\rangle\langle \xi_\nu| \\
    &+ \int^t_0 dt_1\, e^{\mathcal{K}(t-t_1)} \mathcal{L}(t_1) e^{\mathcal{K}t_1}\rho_{\text{sys}}(0)\otimes|\xi_\mu\rangle\langle \xi_\nu| \\
    &+ \int^t_0 dt_2 \int^{t_2}_0 dt_1\, e^{\mathcal{K}(t-t_2)}\mathcal{L}(t_2)e^{\mathcal{K}(t_2-t_1)}\mathcal{L}(t_1)e^{\mathcal{K}t_1}\rho_{\text{sys}}(0)\otimes|\xi_\mu\rangle\langle \xi_\nu|\\
    & + \cdots
\end{split}
\end{align}
Taking the time derivative of Eq. (\ref{Eq:rho_mu_nu_expansion}), we have
\begin{align}
\begin{split}
    \frac{d}{dt}\rho_{\mu,\nu}(t)&= \text{Tr}_{\text{field}}\hat{\mathcal{N}}\Big(\mathcal{K}R_{\mu,\nu}(t)+\mathcal{L}(t)R_{\mu,\nu}(t)\Big) \\
    &= \mathcal{K}\rho_{\mu,\nu}(t) + \text{Tr}_{\text{field}}\hat{\mathcal{N}}\Big(-\big[L^\dagger a(t), R_{\mu,\nu}(t)\big] + \big[L a^\dagger(t), R_{\mu,\nu}(t)\big]\Big) \\
    &=\mathcal{K}\rho_{\mu,\nu}(t) -\Big[L^\dagger, \text{Tr}_{\text{field}}\hat{\mathcal{N}}\big(a(t)R_{\mu,\nu}(t)\big)\Big] + \Big[L , \text{Tr}_{\text{field}}\hat{\mathcal{N}}\big(a^\dagger(t)R_{\mu,\nu}(t)\big)\Big].
\label{Eq:Fock_state_master_integral}
\end{split}
\end{align}
In the third line, we take the system operators $L^\dagger$ and $L$ outside of the partial trace over field. Due to normal ordering, the placement of field operators inside normal ordering is unimportant. Therefore, we use the property $\hat{\mathcal{N}}(a(t)R_{\mu, \nu}(t)) = \hat{\mathcal{N}}(R_{\mu, \nu}(t) a(t))$ to combine the field operators $a(t)$ and $a^\dagger(t)$ with $R_{\mu,\nu}$. 
\par
The system operator $\text{Tr}_{\text{field}}\hat{\mathcal{N}}\big(a(t)R_{\mu,\nu}(t)\big)$ is expanded as
\begin{align}
\begin{split}
    \text{Tr}_{\text{field}}\hat{\mathcal{N}}\big(a(t)R_{\mu,\nu}(t)\big) &=\,\text{Tr}_{\text{field}}\hat{\mathcal{N}}\bigg(e^{\mathcal{K}t}\rho_{\text{sys}}(0)\otimes a(t)|\xi_\mu\rangle\langle \xi_\nu| \\
    &\quad + \int^t_0 dt_1\, e^{\mathcal{K}(t-t_1)} \mathcal{L}(t_1) e^{\mathcal{K}t_1}\rho_{\text{sys}}(0)\otimes a(t)|\xi_\mu\rangle\langle \xi_\nu| \\
    &\quad + \int^t_0 dt_2 \int^{t_2}_0 dt_1\, e^{\mathcal{K}(t-t_2)}\mathcal{L}(t_2)e^{\mathcal{K}(t_2-t_1)}\mathcal{L}(t_1)e^{\mathcal{K}t_1}\rho_{\text{sys}}(0)\otimes a(t)|\xi_\mu\rangle\langle \xi_\nu|\\
    &\quad + \cdots \bigg).
\label{Eq:lower_R_mu_nu_1}
\end{split}
\end{align}
We have moved $a(t)$ to the left of $|\xi_\mu\rangle\langle\xi_\nu|$ due to the normal ordering. 
Using the property 
\begin{equation}
    a(t)|\xi_\mu\rangle = \sqrt{\mu} \xi(t) |\xi_{\mu-1}\rangle,
\end{equation}
we can simplify Eq. (\ref{Eq:lower_R_mu_nu_1}) as
\begin{equation}
    \text{Tr}_{\text{field}}\hat{\mathcal{N}}\big(a(t)R_{\mu,\nu}(t)\big) = \sqrt{\mu}\xi(t)\rho_{\mu-1, \nu}.
\end{equation}
Similarly, $\text{Tr}_{\text{field}}\hat{\mathcal{N}}\big(a^\dagger(t)R_{\mu,\nu}(t)\big) = \sqrt{\nu}\xi^*(t)\rho_{\mu, \nu-1}$. Therefore we can re-write Eq. (\ref{Eq:Fock_state_master_integral}) to obtain the Fock state master equation as
\begin{equation}
    \frac{d}{dt}\rho_{\mu,\nu}(t) = \mathcal{K}\rho_{\mu,\nu}(t) - \sqrt{\mu}\xi(t)[L^\dagger, \rho_{\mu-1,\nu}] + \sqrt{\nu}\xi^*(t) [L, \rho_{\mu,\nu-1}].
\label{Eq:Fock_state_master}
\end{equation}
We note that the Fock state master equations are first derived using the quantum stochastic differential formalism \cite{Baragiola_2012}. They can also be derived by directly taking the time derivative of Eq. (\ref{Eq:def_rho_mu_nu}) using ordinary differential calculus \cite{Ko_2022}.
Eq. (\ref{Eq:Fock_state_master}) is a set of differential equations that couples a hierarchy of auxiliary density matrices $\rho_{\mu,\nu}$, with the highest hierarchy density matrix $\rho_{m,m}$ being the physical reduced system state. The auxiliary density matrices encode non-Markovian effects from the $m$-photon Fock state input \cite{Baragiola_2012,Ko_2022}.

\section{Deriving the master equation under a Markovian Gaussian input}

Mean-zero Markovian Gaussian states are completely characterized by their two-point correlation functions \cite{gardiner_zoller_quantum_noise}:
\begin{subequations}
    \begin{equation}
        \langle a^\dagger(t_2)a(t_1)\rangle = n\delta(t_2-t_1)
    \end{equation}
    \begin{equation}
        \langle a(t_2)a^\dagger(t_1)\rangle = (n+1)\delta(t_2-t_1)
    \end{equation}
    \begin{equation}
        \langle a(t_2)a(t_1)\rangle = m \delta(t_2-t_1)
    \end{equation}
    \begin{equation}
        \langle a^\dagger(t_2)a^\dagger(t_1)\rangle= m^* \delta(t_2-t_1).
    \end{equation}
\label{Eq:two_point_correlations}
\end{subequations}
The higher order correlation functions satisfy the Wick property, such that all odd-point correlation functions are zero, and all higher even-point correlation functions can be factorized in terms of the two-point correlation functions \cite{Rammer_book}, i.e., 
\begin{equation}
    \langle a^{(\dagger)}(t_{2n})\cdots a^{(\dagger)}(t_{1})\rangle = \sum_{\text{a.p.p.}} \prod_{k>l} \langle a^{(\dagger)}(t_{k})a^{(\dagger)}(t_{l}) \rangle,
\label{Eq:Wick_thm}
\end{equation}
where a.p.p. stands for all possible pairs $(k,l)$ from the list $2n, 2n-1, \cdots, 1$. The notation $a^{(\dagger)}(t)$ means either $a^\dagger(t)$ or $a(t)$.
A Markovian squeezed vacuum state and a Markovian thermal state are both examples of mean-zero Markovian Gaussian states.

Since all odd-order correlation functions are zero, the odd-order expansion terms vanish, and the normal-ordered perturbative expansion becomes
\begin{align}
\begin{split}
    \rho_{\text{sys}}(t) = \text{Tr}_{\text{field}}\hat{\mathcal{N}}\Bigg( &e^{\mathcal{K}t}\rho_{\text{tot}}(0) \\
    &+ \int^t_0 dt_2 \int^{t_2}_0 dt_1\, e^{\mathcal{K}(t-t_2)}\mathcal{L}(t_2)e^{\mathcal{K}(t_2-t_1)}\mathcal{L}(t_1)e^{\mathcal{K}t_1}\rho_{\text{tot}}(0)\\
    & + \int^t_0 dt_4 \int^{t_4}_0 dt_3 \int^{t_3}_0 dt_2 \int^{t_2}_0 dt_1 \, e^{\mathcal{K}(t-t_4)}\mathcal{L}(t_4)e^{\mathcal{K}(t_4-t_3)}\mathcal{L}(t_3)\\
    &\qquad\qquad\qquad\qquad e^{\mathcal{K}(t_3-t_2)}\mathcal{L}(t_2)e^{\mathcal{K}(t_2-t_1)}\mathcal{L}(t_1)e^{\mathcal{K}t_1} \rho_{\text{tot}}(0) \\
    &+\cdots \Bigg) \\
    = S_0 + S_2 +& S_4 + \cdots,
\label{Eq:normal_ordered_reduced_system_even_order}
\end{split}
\end{align}
where $S_{2n}$ represents the $2n$-th order term.
To simplify this expression, we note that $S_{2n}$ consists of terms that contain $2n$-point correlation functions and take the general form
\begin{equation}
    \int_{t\geq t_{2n}\geq\cdots\geq t_1\geq 0} dt_{2n} \cdots dt_1 \big\langle :a^{(\dagger)}(t_{2n})\cdots a^{(\dagger)}(t_{1}):\big\rangle \zeta(t_{2n}, \cdots, t_1).
\label{Eq:2n_term_general_form}
\end{equation}
We have used the notation $\langle :A :\rangle$ to denote the normal-ordered expectation value, so that the $a^{(\dagger)}(t)$ are ordered.
$\zeta(t_{2n}, \cdots, t_1)$ is some operator whose explicit form does not concern us right now. According to Wick's theorem (see Eq. (\ref{Eq:Wick_thm}) in the main text), the correlation function can be factorized into two-point correlation functions, which are proportional to the delta function. Due to the delta functions, only the ordered pairing
\begin{equation}
    \big\langle : a^{(\dagger)}(t_{2n})a^{(\dagger)}(t_{2n-1}):\big\rangle \cdots \big\langle :a^{(\dagger)}(t_{2})a^{(\dagger)}(t_{1}):\big\rangle
\end{equation}
out of the $(2n-1)(2n-3)\cdots 1$ possible pairings contribute to the integral of Eq. (\ref{Eq:2n_term_general_form}). All correlation functions in this ordered pairing contain adjacent time points. Any other pairing will have at least one correlation function consisting of non-adjacent time points $(t_i, t_j)$. Because the delta function restricts $t_i$ to be equal to $t_j$, all the time points $t_k$ between $t_i$ and $t_j$ will also be restricted to be equal to $t_i$ due to the time ordering in the integral (see Eq. (\ref{Eq:2n_term_general_form})). Integrating over these intermediate time points $t_k$ over a single point therefore makes the integral in Eq. (\ref{Eq:2n_term_general_form}) equal to $0$.
\par
Using the fact that only the ordered pairing contributes non-trivially to Eq. (\ref{Eq:2n_term_general_form}) and using Eq. (\ref{Eq:two_point_correlations}) in the main text, we can now integrate out the delta function $\delta(t_{2n}-t_{2n-1})$ and write $S_{2n}$ in Eq. (\ref{Eq:normal_ordered_reduced_system_even_order}) as
\begin{align}
\begin{split}
    S_{2n} = &\int_{t\geq t' \geq t_{2n-2}\geq\cdots\geq t_1\geq 0} dt' dt_{2n-2} \cdots dt_1 \, e^{\mathcal{K}(t-t')} \mathcal{J} \\
    & \qquad \text{Tr}_{\text{field}}\hat{\mathcal{N}}\Big(  e^{\mathcal{K}(t'-t_{2n-2})} \mathcal{L}(t_{2n-2}) e^{\mathcal{K}(t_{2n-2}-t_{2n-3})}\mathcal{L}(t_{2n-3}) \cdots e^{\mathcal{K}(t_2-t_1)}\mathcal{L}(t_1)e^{\mathcal{K}t_1}\rho_{\text{tot}}(0)\Big),
\end{split}
\end{align}
where the superoperator $\mathcal{J}$ is equal to
\begin{align}
\begin{split}
    \mathcal{J} =& \frac{m}{2} [L^\dagger, [L^\dagger, \bullet] ] + \frac{m^*}{2} [L, [L, \bullet] ] \\
    &- \frac{n}{2}[L^\dagger, [L, \bullet]] - \frac{n}{2}[L, [L^\dagger, \bullet]].
\end{split}
\end{align}
The factors of $1/2$ originate from the fact the integral $\int^t dt_{2n} \int^{t_{2n}} dt_{2n-1} \delta(t_{2n}-t_{2n-1})$ only integrates over half of the delta function. Taking the time derivative of $S_{2n}$, we have
\begin{equation}
    \frac{d}{dt} S_{2n} = \mathcal{K} S_{2n} + \mathcal{J} S_{2n-2}.
\end{equation}
Substituting this relation into Eq. (\ref{Eq:normal_ordered_reduced_system_even_order}), we obtain the following Markovian master equation \cite{Gardiner1985}
\begin{align}
\begin{split}
    \frac{d}{dt}\rho_{\text{sys}}(t) &=  (\mathcal{K}+\mathcal{J} )\rho_{\text{sys}}(t) \\
    &= -i[H_{\text{sys}}, \rho_{\text{sys}}(t)] \\
    &\quad - m\big(L^\dagger \rho_{\text{sys}}(t) L^\dagger - \frac{1}{2} L^\dagger L^\dagger \rho_{\text{sys}}(t) -  \frac{1}{2} \rho_{\text{sys}}(t) L^\dagger L^\dagger\big) \\
    &\quad -m^*\big(L \rho_{\text{sys}}(t) L - \frac{1}{2}LL\rho_{\text{sys}}(t) - \frac{1}{2}\rho_{\text{sys}}(t) LL\big) \\
    &\quad + n\big(L^\dagger \rho_{\text{sys}}(t) L - \frac{1}{2}L L^\dagger \rho_{\text{sys}}(t) - \frac{1}{2}\rho_{\text{sys}}(t)L L^\dagger\big) \\
    &\quad + (n+1)\big(L\rho_{\text{sys}}(t)L^\dagger - \frac{1}{2}L^\dagger L\rho_{\text{sys}}(t) - \frac{1}{2}\rho_{\text{sys}}(t)L^\dagger L\big).
\end{split}
\end{align}

\section{Shorthand notation to express the perturbative terms}
\begin{figure}[h]
    \centering
    \includegraphics[scale=0.6]{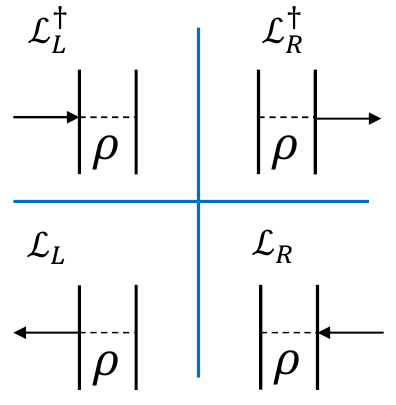}
    \caption{Rules for drawing the four different interactions in the double-sided Feynman diagram. If there are multiple interactions in a pathway, time goes from the bottom to the top.}
    \label{fig:Feynman_diagram_rules}
\end{figure}

\begin{figure}[h]
    \centering
    \includegraphics[scale=0.5]{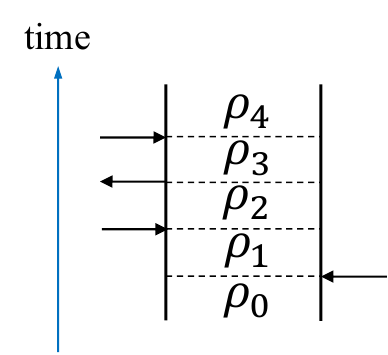}
    \caption{The Feynman diagram corresponding to the pathway $(\mathcal{L}^\dagger_L \mathcal{L}_L \mathcal{L}^\dagger_L \mathcal{L}_R)$.}
    \label{fig:Feynman_diagram_expample}
\end{figure}

In both conventional and normal-ordered perturbative expansions, the interaction superoperator $\mathcal{L}$ can be decomposed as a sum of four types of interactions, i.e., 
\begin{equation}
    \mathcal{L} = \mathcal{L}^\dagger_L + \mathcal{L}^\dagger_R + \mathcal{L}_L + \mathcal{L}_R,
\end{equation}
where we have introduced the following shorthands:
\begin{subequations}
\begin{equation}
    \mathcal{L}^\dagger_L = -a(t)L^\dagger \bullet
\end{equation}    
\begin{equation}
    \mathcal{L}^\dagger_R = \bullet a(t)L^\dagger
\end{equation}
\begin{equation}
    \mathcal{L}_L = a^\dagger(t)L \bullet
\end{equation}
\begin{equation}
    \mathcal{L}_R = - \bullet a^\dagger(t) L.
\end{equation}
\label{Eq:4_L_superoperators}
\end{subequations}
Since a sequence of interactions completely specifies the perturbative term, we can express the lengthy integral expressions succinctly using the interaction pathways $(\cdots \mathcal{L}_3 \mathcal{L}_2 \mathcal{L}_1)$, where $\mathcal{L}_1$ is the first interaction at time $t_1$, $\mathcal{L}_2$ is the next interaction at time $t_2$, and so on. The $n$-th order perturbation contains $4^n$ interaction pathways that need to be summed. However, we will see that many interaction pathway terms are $0$. We note that these interaction pathways are commonly represented by the double-sided Feynman diagrams \cite{Mukamel_book}. The rules for drawing the four different interactions are given in Fig. (\ref{fig:Feynman_diagram_rules}).
For example, the double-sided Feynman diagram of the pathway $(\mathcal{L}^\dagger_L \mathcal{L}_L \mathcal{L}^\dagger_L \mathcal{L}_R)$ is shown in Fig. (\ref{fig:Feynman_diagram_expample}).
\par
For normal-ordered expansion, we will use the notation $(\mathcal{L}^\dagger_L \mathcal{L}_L \mathcal{L}^\dagger_L \mathcal{L}_R)_N$ to represent the 4-th order normal-ordered expansion term (note the subscript $N$). The explicit integral expression of this normal-ordered expansion pathway is
\begin{align}
\begin{split}
    (\mathcal{L}^\dagger_L \mathcal{L}_L \mathcal{L}^\dagger_L \mathcal{L}_R)_N = &\int^t_0 dt_4 \int^{t_4}_0 dt_3 \int^{t_3}_0 dt_2 \int^{t_2}_0 dt_1 \\ &e^{\mathcal{K}(t-t_4)}(-L^\dagger \bullet)e^{\mathcal{K}(t_4-t_3)}(L \bullet)e^{\mathcal{K}(t_3-t_2)}(-L^\dagger \bullet)e^{\mathcal{K}(t_2-t_1)}(-\bullet L)e^{\mathcal{K}t_1} \\
    &\big\langle a^\dagger(t_3)a^\dagger(t_1)a(t_4)a(t_2) \big\rangle.
\end{split}
\end{align}
Note that the field correlation function has been normal-ordered. 
\par
Similarly, the corresponding conventional expansion term (note the subscript $C$) has the integral expression
\begin{align}
\begin{split}
    (\mathcal{L}^\dagger_L \mathcal{L}_L \mathcal{L}^\dagger_L \mathcal{L}_R)_C = &\int^t_0 dt_4 \int^{t_4}_0 dt_3 \int^{t_3}_0 dt_2 \int^{t_2}_0 dt_1 \\ &e^{\mathcal{K}'(t-t_4)}(-L^\dagger \bullet)e^{\mathcal{K}'(t_4-t_3)}(L \bullet)e^{\mathcal{K}'(t_3-t_2)}(-L^\dagger \bullet)e^{\mathcal{K}'(t_2-t_1)}(-\bullet L)e^{\mathcal{K}'t_1} \\
    &\big\langle a^\dagger(t_1)a(t_4)a^\dagger(t_3)a(t_2) \big\rangle.
\label{Eq:conventional_sequence_example}
\end{split}
\end{align}
In the conventional expansion term, we have used the free evolution generator $\mathcal{K}'$. The field correlation function is now ordered according to the interaction pathway. The interaction pathway $(\mathcal{L}^\dagger_L \mathcal{L}_L \mathcal{L}^\dagger_L \mathcal{L}_R)_C$ gives rise to the expectation value $\text{Tr}(a(t_4)a^\dagger(t_3)a(t_2)\rho_{\text{field}}(0)a^\dagger(t_1))$, which is equal to the correlation function in Eq. (\ref{Eq:conventional_sequence_example}) after cyclic permutation. Note that this field correlation function is not normal-ordered. 
\par
In the normal-ordered expansion, the field correlation function only depends on whether the interaction is due to $\mathcal{L}$ or $\mathcal{L}^\dagger$. It does not depend on whether the interaction is left-multiplying or right-multiplying. For example, both pathways $(\mathcal{L}^\dagger_L \mathcal{L}_L \mathcal{L}^\dagger_L \mathcal{L}_R)_N$ and $(\mathcal{L}^\dagger_L \mathcal{L}_R \mathcal{L}^\dagger_R \mathcal{L}_R)_N$ have the same correlation function $\langle a^\dagger(t_3)a^\dagger(t_1)a(t_4)a(t_2) \rangle$. These two pathways have the same sequence of $\mathcal{L}$ and $\mathcal{L}^\dagger$ interactions, but differ in their sequences of left- (subscript $L$) and right- (subscript $R$) interactions. Therefore, different pathways in the normal-ordered expansion can be combined, reducing the total number of expansion terms. On the other hand, the corresponding conventional expansion terms have different field correlation functions. The pathway $(\mathcal{L}^\dagger_L \mathcal{L}_L \mathcal{L}^\dagger_L \mathcal{L}_R)_C$ has the correlation function $\langle a^\dagger(t_1)a(t_4)a^\dagger(t_3)a(t_2)\rangle$., while the pathway $(\mathcal{L}^\dagger_L \mathcal{L}_R \mathcal{L}^\dagger_R \mathcal{L}_R)_C$ has the correlation fucntion $\langle a^\dagger(t_1)a(t_2)a^\dagger(t_3)a(t_4)\rangle$. Therefore, in the conventional expansion, different pathways need to be considered separately in general. Furthermore, in order to evaluate the field correlation functions, the correlation functions need to be re-expressed as sums of normal-ordered correlation functions, further increasing the number of expansion terms.
\par
Finally, we note that interaction pathways often appear in Hermitian conjugate pairs, allowing us to reduce the number of pathways we need to calculate by a factor of 2. The superoperators defined in Eq. (\ref{Eq:4_L_superoperators}) have the Hermitian conjugate properties:
\begin{subequations}
\begin{equation}
    (\mathcal{L}_R \rho)^\dagger = \mathcal{L}^\dagger_L \rho^\dagger
\end{equation}
and
\begin{equation}
    (\mathcal{L}_L\rho)^\dagger = \mathcal{L}_R^\dagger \rho^\dagger .
\end{equation}
\end{subequations}
We remind the reader that $\mathcal{L}_L^\dagger$, $\mathcal{L}_R^\dagger$, $\mathcal{L}_L$, and $\mathcal{L}_R$ are superoperators that act on the space of operators, and $\rho$ is an operator that acts on the space of vectors in the Hilbert space. In both the conventional and the normal-ordered expansions, the free evolution generators (which are superoperators) satisfy the properties:
\begin{subequations}
\begin{equation}
    (\mathcal{K}'\rho)^\dagger = \mathcal{K}' \rho^\dagger
\end{equation}
and
\begin{equation}
    (\mathcal{K}\rho)^\dagger = \mathcal{K}\rho^\dagger.
\end{equation}
\end{subequations}
Exponentiating these properties, we have
\begin{subequations}
\begin{equation}
    (e^{\mathcal{K}'t} \rho)^\dagger = e^{\mathcal{K}'t} \rho^\dagger
\end{equation}
and
\begin{equation}
    (e^{\mathcal{K}t} \rho)^\dagger = e^{\mathcal{K}t} \rho^\dagger.
\end{equation}
\end{subequations}
Therefore, for example, the pathways $(\mathcal{L}^\dagger_L \mathcal{L}_L \mathcal{L}^\dagger_L \mathcal{L}_R)_{C \text{or}N}$ and $(\mathcal{L}_R \mathcal{L}_R^\dagger \mathcal{L}_R \mathcal{L}^\dagger_L )_{C \text{or}N}$ are Hermitian conjugates of each other. In terms of the double-sided Feynman diagrams, two diagrams that are reflections (around the middle of the ladder) of each other are Hermitian conjugates of each other.

\section{Normal-ordered perturbative expansion terms}
For normal-ordered expansion, the correlation functions in the interaction pathways are independent of the interaction direction (i.e., left- or right-multiplying). Therefore, we can combine the left- and right-interactions and define
\begin{subequations}
\begin{equation}
    \Lambda^\dagger = [-a(t) L^\dagger , \bullet]
\end{equation}
\begin{equation}
    \Lambda = [a^\dagger(t)L, \bullet],
\end{equation}
\end{subequations}
such that $\mathcal{L}$ is decomposed as a sum of two types of interactions, i.e., 
\begin{equation}
    \mathcal{L} = \Lambda^\dagger + \Lambda.
\end{equation}
Note that $(\Lambda\rho)^\dagger = \Lambda^\dagger\rho^\dagger$
Now, the $n$-th order normal-ordered perturbation only contains $2^n$ pathways. 
\par
\begin{figure}[h]
    \centering
    \includegraphics[scale=0.6]{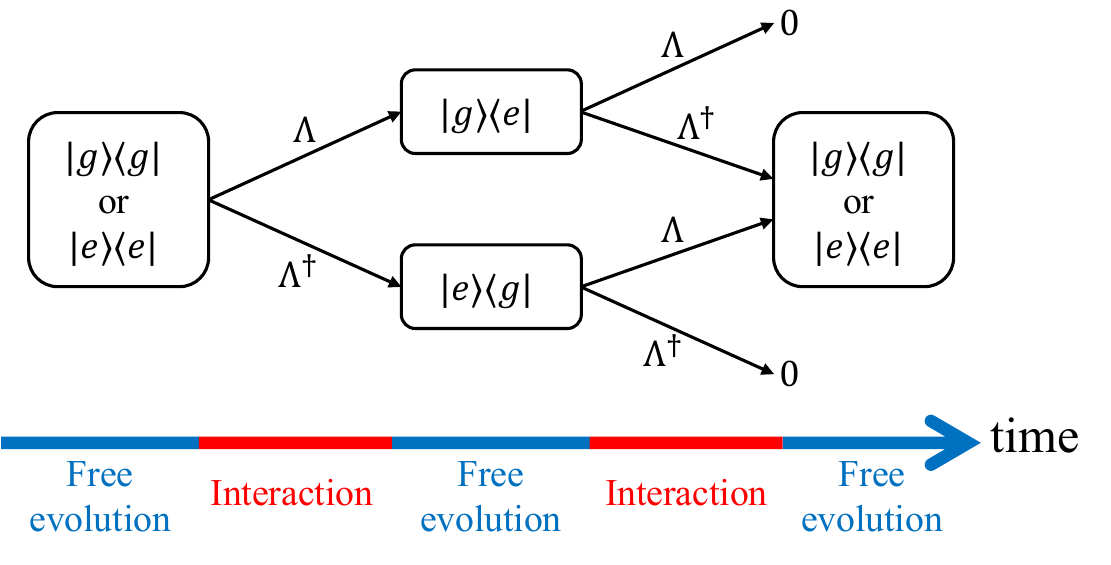}
    \caption{If the initial density matrix is nonzero only in the population elements, then the first two interactions have to be either $(\Lambda\Lambda^\dagger)_N$ or $(\Lambda^\dagger \Lambda)_N$. The same argument applies to the third and fourth interactions, or the fifth and sixth interactions, and so on.}
    \label{fig:normal_order_nonzero_terms}
\end{figure}

We can further reduce the number of pathways we need to consider by examining how the density matrix elements evolve in different pathways.
In our model of the two level system, the free evolution can mix the population elements (i.e., $|g\rangle\langle g|$ and $|e\rangle\langle e|$), but it does not mix between the population elements and the off-diagonal coherence elements (i.e., $|g\rangle\langle e|$ and $|e\rangle\langle g|$). Different coherence elements also don't mix with each other. The initial state of our system is $|g\rangle\langle g|$, which is a population element. Applying $\Lambda$ to the population elements always results in $|g\rangle \langle e|$. Applying $\Lambda$ again to $|g\rangle \langle e|$ results in $0$, but applying $\Lambda^\dagger$ to $|g\rangle \langle e|$ results in the population elements. Similarly, if we first apply $\Lambda^\dagger$ to the initial state, the next interaction has to be $\Lambda$ for the pathway to be nonzero. These pathways are illustrated in Fig. (\ref{fig:normal_order_nonzero_terms}). Therefore, the first two interactions have to be either $(\Lambda\Lambda^\dagger)_N$ or $(\Lambda^\dagger \Lambda)_N$. Similarly, the third and the fourth interactions also have to be either $(\Lambda\Lambda^\dagger)_N$ or $(\Lambda^\dagger \Lambda)_N$. The same argument applies to every pair of interactions afterwards. Hence, the $n$-th order ($n$ is even) normal-ordered perturbation contains $2^{n/2}$ nonzero pathways. 
\par
Since we are interested in the excited state population $|e\rangle\langle e|$, which only occurs in even-order perturbations, we will only compute the even-order terms up to the 6\textsuperscript{th} order.

\subsection{2\textsuperscript{nd} order terms in the normal-ordered expansion}
The 2 pathways in the second order normal-ordered expansion are:
\begin{equation*}
    \text{pathway 1: } (\Lambda\Lambda^\dagger)_N 
\end{equation*}
\begin{equation*}
    \text{pathway 2: } (\Lambda^\dagger \Lambda)_N ,
\end{equation*}
which are Hermitian conjugates of each other. Therefore, we only need to compute one pathway.

\subsection{4\textsuperscript{th} order terms in the normal-ordered expansion}
The 4 pathways in the fourth order normal-ordered expansion are:
\begin{equation*}
    \text{pathway 1: } (\Lambda\Lambda^\dagger\Lambda\Lambda^\dagger)_N 
\end{equation*}
\begin{equation*}
    \text{pathway 2: } (\Lambda\Lambda^\dagger\Lambda^\dagger \Lambda)_N 
\end{equation*}
\begin{equation*}
    \text{pathway 3: } (\Lambda^\dagger \Lambda\Lambda^\dagger \Lambda)_N
\end{equation*}
\begin{equation*}
    \text{pathway 4: } (\Lambda^\dagger \Lambda\Lambda\Lambda^\dagger)_N 
\end{equation*}
Pathways 1 and 2 are Hermitian conjugates of pathways 3 and 4, respectively. Therefore, we only need to compute two pathways.

\subsection{6\textsuperscript{th} order terms in the normal-ordered expansion}
The 8 pathways in the sixth order normal-ordered expansion are:
\begin{equation*}
    \text{pathway 1: } (\Lambda\Lambda^\dagger\Lambda\Lambda^\dagger\Lambda\Lambda^\dagger)_N 
\end{equation*}
\begin{equation*}
    \text{pathway 2: } (\Lambda\Lambda^\dagger\Lambda\Lambda^\dagger\Lambda^\dagger \Lambda)_N 
\end{equation*}
\begin{equation*}
    \text{pathway 3: } (\Lambda\Lambda^\dagger\Lambda^\dagger \Lambda\Lambda\Lambda^\dagger)_N 
\end{equation*}
\begin{equation*}
    \text{pathway 4: } (\Lambda\Lambda^\dagger\Lambda^\dagger \Lambda\Lambda^\dagger \Lambda)_N 
\end{equation*}
\begin{equation*}
    \text{pathway 5: } (\Lambda^\dagger \Lambda\Lambda^\dagger \Lambda\Lambda^\dagger \Lambda)_N  
\end{equation*}
\begin{equation*}
    \text{pathway 6: } (\Lambda^\dagger \Lambda\Lambda^\dagger \Lambda\Lambda\Lambda^\dagger)_N  
\end{equation*}
\begin{equation*}
    \text{pathway 7: } (\Lambda^\dagger \Lambda\Lambda\Lambda^\dagger\Lambda^\dagger \Lambda)_N 
\end{equation*}
\begin{equation*}
    \text{pathway 8: } (\Lambda^\dagger \Lambda\Lambda\Lambda^\dagger\Lambda\Lambda^\dagger)_N 
\end{equation*}
Pathways 1 to 4 are Hermitian conjugates of pathways 5 to 8. Therefore, we only need to compute four pathways.

\section{Conventional perturbative expansion terms}
Since the field correlation functions are most easily evaluated in the normal-ordered form, we will re-express the non-normal-ordered correlation functions in the conventional expansion in terms of normal-ordered correlation functions. To do so, we introduce some new notations to represent the integral expressions compactly.
\par
We define the interaction superoperators
\begin{subequations}
\begin{equation}
    \mathcal{S}_L = -L^\dagger L \bullet
\end{equation}
\begin{equation}
    \mathcal{S}_R = -\bullet L^\dagger L
\end{equation}
\begin{equation}
    \mathcal{S}_M = L \bullet L^\dagger.
\end{equation}
\end{subequations}
For example, the correlation function in Eq. (\ref{Eq:conventional_sequence_example}) can be re-expressed in terms of normal-ordered correlation functions as
\begin{equation}
    \big\langle a^\dagger(t_1)a(t_4)a^\dagger(t_3)a(t_2) \big\rangle = \big\langle a^\dagger(t_1)a^\dagger(t_3)a(t_4)a(t_2) \big\rangle + \delta(t_4-t_3)\big\langle a^\dagger(t_1)a(t_2)\big\rangle.
\end{equation}
Re-indexing the time variables, the entire integral expression of Eq. (\ref{Eq:conventional_sequence_example}) can then be written as
\begin{align}
\begin{split}
    (\mathcal{L}^\dagger_L \mathcal{L}_L \mathcal{L}^\dagger_L \mathcal{L}_R)_C = &\int^t_0 dt_4 \int^{t_4}_0 dt_3 \int^{t_3}_0 dt_2 \int^{t_2}_0 dt_1 \\ &e^{\mathcal{K}'(t-t_4)}(-L^\dagger \bullet)e^{\mathcal{K}'(t_4-t_3)}(L \bullet)e^{\mathcal{K}'(t_3-t_2)}(-L^\dagger \bullet)e^{\mathcal{K}'(t_2-t_1)}(-\bullet L)e^{\mathcal{K}'t_1} \\
    &\big\langle a^\dagger(t_1)a(t_4)a^\dagger(t_3)a(t_2) \big\rangle \\
    = & \int^t_0 dt_4 \int^{t_4}_0 dt_3 \int^{t_3}_0 dt_2 \int^{t_2}_0 dt_1 \\ &e^{\mathcal{K}'(t-t_4)}(-L^\dagger \bullet)e^{\mathcal{K}'(t_4-t_3)}(L \bullet)e^{\mathcal{K}'(t_3-t_2)}(-L^\dagger \bullet)e^{\mathcal{K}'(t_2-t_1)}(-\bullet L)e^{\mathcal{K}'t_1} \\
    &\big\langle a^\dagger(t_1)a^\dagger(t_3)a(t_4)a(t_2) \big\rangle \\
    & + \frac{1}{2} \int^t_0 dt_3 \int^{t_3}_0 dt_2 \int^{t_2}_0 dt_1 \\ 
    & e^{\mathcal{K}'(t-t_3)}(-L^\dagger L \bullet)e^{\mathcal{K}'(t_3-t_2)}(-L^\dagger \bullet)e^{\mathcal{K}'(t_2-t_1)}(-\bullet L)e^{\mathcal{K}'t_1} \\
    &\big\langle a^\dagger(t_1)a(t_2) \big\rangle \\ 
    = &\, (:\mathcal{L}^\dagger_L \mathcal{L}_L \mathcal{L}^\dagger_L \mathcal{L}_R:)_C + \frac{1}{2}(:\mathcal{S}_L \mathcal{L}^\dagger_L \mathcal{L}_R:)_C.
\end{split}
\end{align}
In the last line, we use the $::$ notation in the conventional expansion pathway $(:\mathcal{L}_n\cdots\mathcal{L}_1:)_C$ to mean that the field correlation functions are normal-ordered, while the free evolution is generated by $\mathcal{K}'$. Note that in the normal-ordered expansion, the free evolution is generated by $\mathcal{K}$. The factor of $1/2$ comes from integrating half of the delta function $\delta(t_4-t_3)$.

\subsection{2\textsuperscript{nd} order terms in the conventional expansion}
\begin{figure}[h]
    \centering
    \includegraphics[scale=0.7]{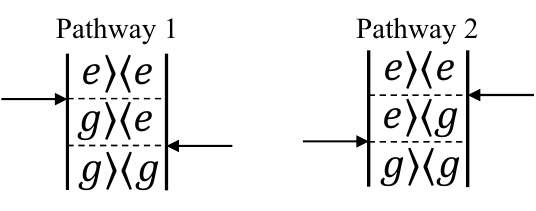}
    \caption{Double-sided Feynman diagrams of the second order pathways in the conventional expansion.}
    \label{fig:conv_2nd}
\end{figure}

There are 2 second order pathways that contribute to the excited state population $|e\rangle\langle e|$. They are
\begin{equation*}
    \text{pathway 1: } (\mathcal{L}^\dagger_L\mathcal{L}_R)_C = (:\mathcal{L}^\dagger_L\mathcal{L}_R:)_C
\end{equation*}
\begin{equation*}
    \text{pathway 2: } (\mathcal{L}_R \mathcal{L}^\dagger_L)_C = (:\mathcal{L}_R \mathcal{L}^\dagger_L:)_C 
\end{equation*}
The two pathways are Hermitian conjugate conjugates of each other. The field correlation functions are already normal-ordered.
Fig. (\ref{fig:conv_2nd}) depicts the double-sided diagrams for these two second order pathways in the conventional expansion.

\subsection{4\textsuperscript{th} order terms in the conventional expansion}
\begin{figure}[h]
    \centering
    \includegraphics[scale=0.6]{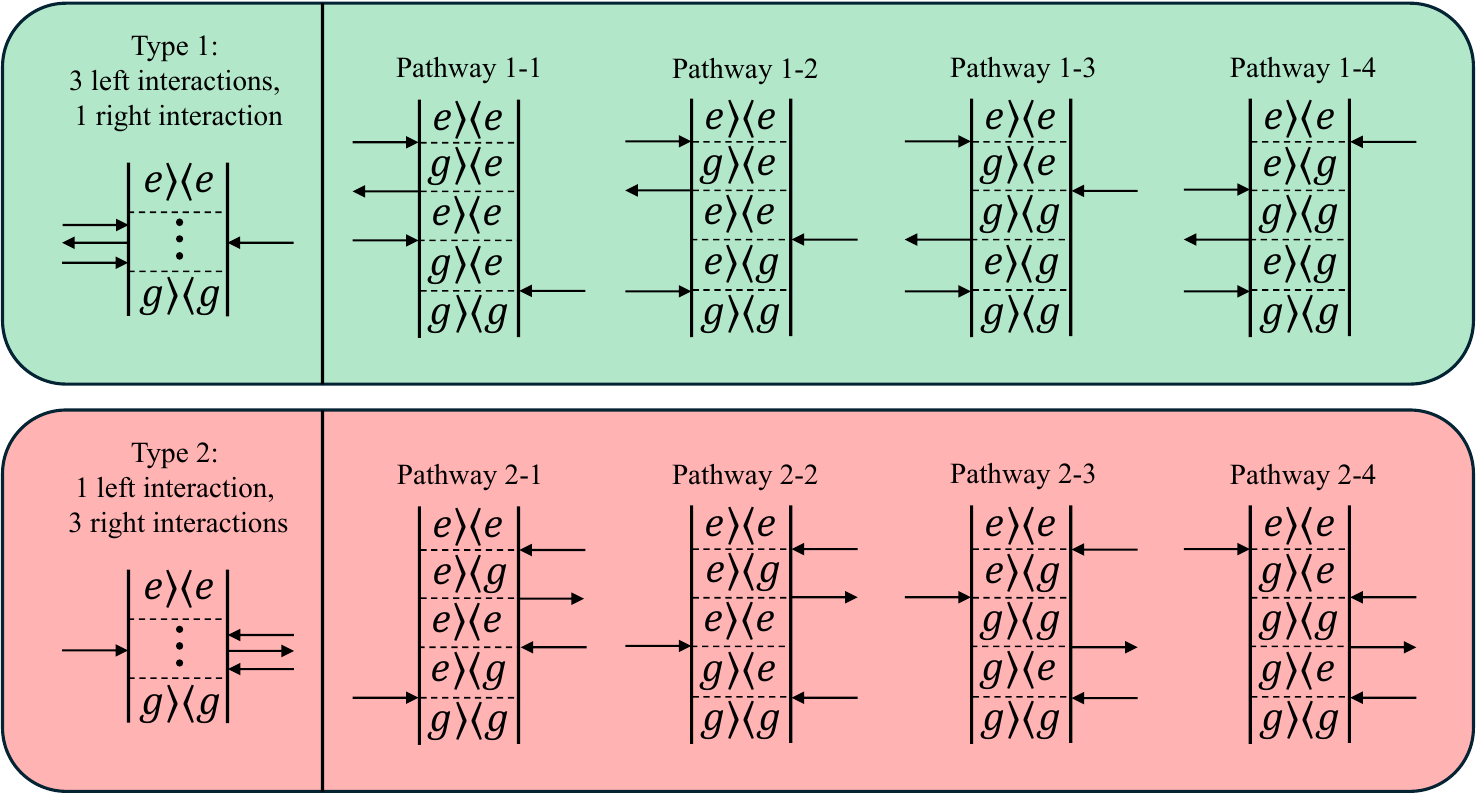}
    \caption{Double-sided Feynman diagrams of the fourth order pathways in the conventional expansion.}
    \label{fig:conv_4th}
\end{figure}

We classify the 4\textsuperscript{th} order terms into two types. The first type of pathway has 3 left interactions and 1 right interaction. The second type of pathway has 1 left interaction and 3 right interactions. The double-sided Feynman diagrams are shown in Fig. (\ref{fig:conv_4th}). Type 2 pathways are reflections of Type 1 pathways, so they are Hermitian conjugates of Type 1 pathways. Therefore, we will only show the expressions for Type 1 pathways.
\par
The expressions for Type 1 pathways are
\begin{align*}
\begin{split}
    &\text{pathway 1-1: } (\mathcal{L}^\dagger_L\mathcal{L}_L\mathcal{L}^\dagger_L\mathcal{L}_R )_C = (:\mathcal{L}^\dagger_L \mathcal{L}_L \mathcal{L}^\dagger_L \mathcal{L}_R:)_C + \frac{1}{2}(:\mathcal{S}_L \mathcal{L}^\dagger_L \mathcal{L}_R:)_C \\
    &\text{pathway 1-2: } (\mathcal{L}^\dagger_L\mathcal{L}_L\mathcal{L}_R\mathcal{L}^\dagger_L )_C = (:\mathcal{L}^\dagger_L\mathcal{L}_L\mathcal{L}_R\mathcal{L}^\dagger_L:)_C + \frac{1}{2}(:\mathcal{S}_L \mathcal{L}_R\mathcal{L}^\dagger_L :)_C \\
    &\text{pathway 1-3: } (\mathcal{L}^\dagger_L\mathcal{L}_R\mathcal{L}_L\mathcal{L}^\dagger_L )_C = (:\mathcal{L}^\dagger_L\mathcal{L}_R\mathcal{L}_L\mathcal{L}^\dagger_L:)_C \\
    &\text{pathway 1-4: } (\mathcal{L}_R\mathcal{L}^\dagger_L\mathcal{L}_L\mathcal{L}^\dagger_L )_C = (:\mathcal{L}_R\mathcal{L}^\dagger_L\mathcal{L}_L\mathcal{L}^\dagger_L:)_C + \frac{1}{2}(:\mathcal{L}_R\mathcal{S}_L \mathcal{L}^\dagger_L :)_C 
\end{split}
\end{align*}

\subsection{6\textsuperscript{th} order terms in the conventional expansion}
\begin{figure}[h]
    \centering
    \includegraphics[scale=0.53]{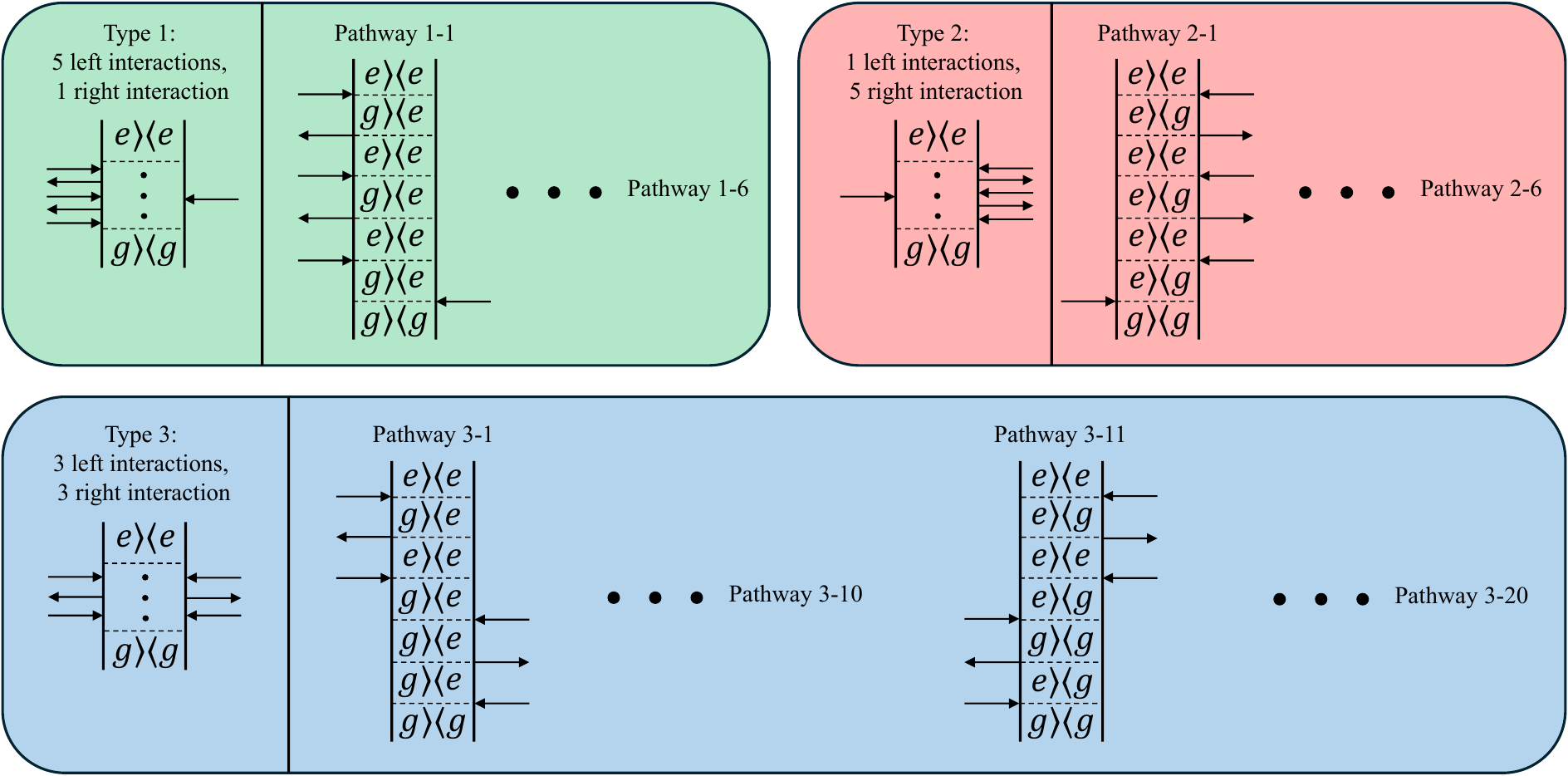}
    \caption{Double-sided Feynman diagrams of the sixth order pathways in the conventional expansion.}
    \label{fig:conv_6th}
\end{figure}

We classify the 6\textsuperscript{th} order terms into three types. Type 1 pathways have 5 left interactions and 1 interaction. Type 2 pathways have 1 left interaction and 5 right interactions. Type 2 pathways and Type 1 pathways are Hermitian conjugates of each other. Type 3 pathways have 3 left interactions and 3 right interactions. There are $\frac{6!}{3!3!}=20$ pathways in Type 3, and 10 of them are Hermitian conjugates of the other 10. We let pathways 3-1 to 3-10 have the first interaction being $\mathcal{L}_R$, and pathways 3-11 to 3-20 have the first interaction being $\mathcal{L}^\dagger_L$. Therefore, pathways 3-11 to 3-20 are the Hermitian conjugates of pathways 3-1 to 3-10. Fig. (\ref{fig:conv_6th}) shows the double-sided Feynman diagrams of the three types of sixth order conventional expansion pathways.
We will only write down the expressions for pathways 1-1 to 1-6 and pathways 3-1 to 3-10. The remaining pathways are the Hermitian conjugates of these pathways.
\par
The Type 1 pathways are
\begin{align*}
    \text{pathway 1-1: } (\mathcal{L}^\dagger_L \mathcal{L}_L \mathcal{L}^\dagger_L \mathcal{L}_L \mathcal{L}^\dagger_L \mathcal{L}_R)_C = &(:\mathcal{L}^\dagger_L \mathcal{L}_L \mathcal{L}^\dagger_L \mathcal{L}_L \mathcal{L}^\dagger_L \mathcal{L}_R:)_C + \frac{1}{2}(:\mathcal{S}_L\mathcal{L}^\dagger_L \mathcal{L}_L \mathcal{L}^\dagger_L \mathcal{L}_R:)_C \\
    &+ \frac{1}{2}(:\mathcal{L}^\dagger_L \mathcal{L}_L \mathcal{S}_L \mathcal{L}^\dagger_L \mathcal{L}_R:)_C + \frac{1}{4}(:\mathcal{S}_L\mathcal{S}_L\mathcal{L}^\dagger_L \mathcal{L}_R:)_C 
\end{align*}
\begin{align*}
    \text{pathway 1-2: } (\mathcal{L}^\dagger_L \mathcal{L}_L \mathcal{L}^\dagger_L \mathcal{L}_L \mathcal{L}_R \mathcal{L}^\dagger_L)_C = &(:\mathcal{L}^\dagger_L \mathcal{L}_L \mathcal{L}^\dagger_L \mathcal{L}_L \mathcal{L}_R \mathcal{L}^\dagger_L:)_C + \frac{1}{2}(:\mathcal{S}_L \mathcal{L}^\dagger_L \mathcal{L}_L \mathcal{L}_R \mathcal{L}^\dagger_L:)_C \\
    &+ \frac{1}{2} (:\mathcal{L}^\dagger_L \mathcal{L}_L \mathcal{S}_L \mathcal{L}_R \mathcal{L}^\dagger_L:)_C + \frac{1}{4}(:\mathcal{S}_L \mathcal{S}_L \mathcal{L}_R \mathcal{L}^\dagger_L:)_C 
\end{align*}
\begin{align*}
    \text{pathway 1-3: } (\mathcal{L}^\dagger_L \mathcal{L}_L \mathcal{L}^\dagger_L \mathcal{L}_R \mathcal{L}_L \mathcal{L}^\dagger_L)_C = &(:\mathcal{L}^\dagger_L \mathcal{L}_L \mathcal{L}^\dagger_L \mathcal{L}_R \mathcal{L}_L \mathcal{L}^\dagger_L:)_C + \frac{1}{2}(:\mathcal{S}_L \mathcal{L}^\dagger_L \mathcal{L}_R \mathcal{L}_L \mathcal{L}^\dagger_L:)_C 
\end{align*}
\begin{align*}
    \text{pathway 1-4: } (\mathcal{L}^\dagger_L \mathcal{L}_L \mathcal{L}_R \mathcal{L}^\dagger_L \mathcal{L}_L \mathcal{L}^\dagger_L)_C = &(:\mathcal{L}^\dagger_L \mathcal{L}_L \mathcal{L}_R \mathcal{L}^\dagger_L \mathcal{L}_L \mathcal{L}^\dagger_L:)_C + \frac{1}{2}(:\mathcal{S}_L \mathcal{L}_R \mathcal{L}^\dagger_L \mathcal{L}_L \mathcal{L}^\dagger_L:)_C \\
    &+ \frac{1}{2} (:\mathcal{L}^\dagger_L \mathcal{L}_L \mathcal{L}_R \mathcal{S}_L \mathcal{L}^\dagger_L:)_C + \frac{1}{4}(:\mathcal{S}_L \mathcal{L}_R \mathcal{S}_L \mathcal{L}^\dagger_L:)_C
\end{align*}
\begin{align*}
    \text{pathway 1-5: } (\mathcal{L}^\dagger_L \mathcal{L}_R \mathcal{L}_L \mathcal{L}^\dagger_L \mathcal{L}_L \mathcal{L}^\dagger_L)_C = &(:\mathcal{L}^\dagger_L \mathcal{L}_R \mathcal{L}_L \mathcal{L}^\dagger_L \mathcal{L}_L \mathcal{L}^\dagger_L:)_C + \frac{1}{2}(:\mathcal{L}^\dagger_L \mathcal{L}_R \mathcal{L}_L \mathcal{S}_L \mathcal{L}^\dagger_L:)_C 
\end{align*}
\begin{align*}
    \text{pathway 1-6: } (\mathcal{L}_R \mathcal{L}^\dagger_L \mathcal{L}_L \mathcal{L}^\dagger_L \mathcal{L}_L \mathcal{L}^\dagger_L)_C = &(:\mathcal{L}_R \mathcal{L}^\dagger_L \mathcal{L}_L \mathcal{L}^\dagger_L \mathcal{L}_L \mathcal{L}^\dagger_L:)_C + \frac{1}{2}(:\mathcal{L}_R \mathcal{S}_L \mathcal{L}^\dagger_L \mathcal{L}_L \mathcal{L}^\dagger_L:)_C \\
    &+ \frac{1}{2} (:\mathcal{L}_R \mathcal{L}^\dagger_L \mathcal{L}_L \mathcal{S}_L \mathcal{L}^\dagger_L:)_C + \frac{1}{4}(:\mathcal{L}_R \mathcal{S}_L \mathcal{S}_L \mathcal{L}^\dagger_L:)_C
\end{align*}
\par
The expressions for the Type 3 pathways 3-1 to 3-10 are
\begin{align*}
    \text{pathway 3-1: } (\mathcal{L}^\dagger_L \mathcal{L}_L \mathcal{L}^\dagger_L \mathcal{L}_R \mathcal{L}^\dagger_R \mathcal{L}_R)_C = &(:\mathcal{L}^\dagger_L \mathcal{L}_L \mathcal{L}^\dagger_L \mathcal{L}_R \mathcal{L}^\dagger_R \mathcal{L}_R:)_C + \frac{1}{2}(:\mathcal{S}_L \mathcal{L}^\dagger_L \mathcal{L}_R \mathcal{L}^\dagger_R \mathcal{L}_R:)_C \\
    &+ \frac{1}{2} (:\mathcal{L}^\dagger_L \mathcal{L}_L \mathcal{L}^\dagger_L \mathcal{S}_R \mathcal{L}_R:)_C + \frac{1}{4}(:\mathcal{S}_L \mathcal{L}^\dagger_L \mathcal{S}_R \mathcal{L}_R:)_C
\end{align*}
\begin{align*}
    \text{pathway 3-2: } (\mathcal{L}^\dagger_L \mathcal{L}_L \mathcal{L}_R \mathcal{L}^\dagger_L \mathcal{L}^\dagger_R \mathcal{L}_R)_C = &(:\mathcal{L}^\dagger_L \mathcal{L}_L \mathcal{L}_R \mathcal{L}^\dagger_L \mathcal{L}^\dagger_R \mathcal{L}_R:)_C + \frac{1}{2}(:\mathcal{S}_L \mathcal{L}_R \mathcal{L}^\dagger_L \mathcal{L}^\dagger_R \mathcal{L}_R:)_C 
\end{align*}
\begin{align*}
    \text{pathway 3-3: } (\mathcal{L}^\dagger_L \mathcal{L}_R \mathcal{L}_L \mathcal{L}^\dagger_L \mathcal{L}^\dagger_R \mathcal{L}_R)_C = &(:\mathcal{L}^\dagger_L \mathcal{L}_R \mathcal{L}_L \mathcal{L}^\dagger_L \mathcal{L}^\dagger_R \mathcal{L}_R:)_C 
\end{align*}
\begin{align*}
    \text{pathway 3-4: } (\mathcal{L}_R \mathcal{L}^\dagger_L \mathcal{L}_L \mathcal{L}^\dagger_L \mathcal{L}^\dagger_R \mathcal{L}_R)_C = &(:\mathcal{L}_R \mathcal{L}^\dagger_L \mathcal{L}_L \mathcal{L}^\dagger_L \mathcal{L}^\dagger_R \mathcal{L}_R:)_C + \frac{1}{2}(:\mathcal{L}_R \mathcal{S}_L \mathcal{L}^\dagger_L \mathcal{L}^\dagger_R \mathcal{L}_R:)_C 
\end{align*}
\begin{align*}
    \text{pathway 3-5: } (\mathcal{L}^\dagger_L \mathcal{L}_L \mathcal{L}_R \mathcal{L}^\dagger_R \mathcal{L}^\dagger_L \mathcal{L}_R)_C = &(:\mathcal{L}^\dagger_L \mathcal{L}_L \mathcal{L}_R \mathcal{L}^\dagger_R \mathcal{L}^\dagger_L \mathcal{L}_R:)_C + \frac{1}{2}(:\mathcal{S}_L \mathcal{L}_R \mathcal{L}^\dagger_R \mathcal{L}^\dagger_L \mathcal{L}_R:)_C \\
    &+ \frac{1}{2} (:\mathcal{L}^\dagger_L \mathcal{L}_L \mathcal{S}_R \mathcal{L}^\dagger_L \mathcal{L}_R:)_C + \frac{1}{4}(:\mathcal{S}_L \mathcal{S}_R \mathcal{L}^\dagger_L \mathcal{L}_R:)_C
\end{align*}
\begin{align*}
    \text{pathway 3-6: } (\mathcal{L}^\dagger_L \mathcal{L}_R \mathcal{L}_L \mathcal{L}^\dagger_R \mathcal{L}^\dagger_L \mathcal{L}_R)_C = &(:\mathcal{L}^\dagger_L \mathcal{L}_R \mathcal{L}_L \mathcal{L}^\dagger_R \mathcal{L}^\dagger_L \mathcal{L}_R:)_C + \frac{1}{2}(:\mathcal{L}^\dagger_L \mathcal{L}_R \mathcal{S}_M \mathcal{L}^\dagger_L \mathcal{L}_R:)_C 
\end{align*}
\begin{align*}
    \text{pathway 3-7: } (\mathcal{L}_R \mathcal{L}^\dagger_L \mathcal{L}_L \mathcal{L}^\dagger_R \mathcal{L}^\dagger_L \mathcal{L}_R)_C = &(:\mathcal{L}_R \mathcal{L}^\dagger_L \mathcal{L}_L \mathcal{L}^\dagger_R \mathcal{L}^\dagger_L \mathcal{L}_R:)_C + \frac{1}{2}(:\mathcal{L}_R \mathcal{S}_L \mathcal{L}^\dagger_R \mathcal{L}^\dagger_L \mathcal{L}_R:)_C \\
    &+ \frac{1}{2} (:\mathcal{L}_R \mathcal{L}^\dagger_L \mathcal{S}_M \mathcal{L}^\dagger_L \mathcal{L}_R:)_C
\end{align*}
\begin{align*}
    \text{pathway 3-8: } (\mathcal{L}^\dagger_L \mathcal{L}_R \mathcal{L}^\dagger_R \mathcal{L}_L \mathcal{L}^\dagger_L \mathcal{L}_R)_C = &(:\mathcal{L}^\dagger_L \mathcal{L}_R \mathcal{L}^\dagger_R \mathcal{L}_L \mathcal{L}^\dagger_L \mathcal{L}_R:)_C + \frac{1}{2}(:\mathcal{L}^\dagger_L \mathcal{S}_R \mathcal{L}_L \mathcal{L}^\dagger_L \mathcal{L}_R:)_C \\
    &+ \frac{1}{2} (:\mathcal{L}^\dagger_L \mathcal{L}_R \mathcal{S}_M \mathcal{L}^\dagger_L \mathcal{L}_R:)_C
\end{align*}
\begin{align*}
    \text{pathway 3-9: } (\mathcal{L}_R \mathcal{L}^\dagger_L \mathcal{L}^\dagger_R \mathcal{L}_L \mathcal{L}^\dagger_L \mathcal{L}_R)_C = &(:\mathcal{L}_R \mathcal{L}^\dagger_L \mathcal{L}^\dagger_R \mathcal{L}_L \mathcal{L}^\dagger_L \mathcal{L}_R:)_C + \frac{1}{2}(:\mathcal{L}_R \mathcal{L}^\dagger_L \mathcal{S}_M \mathcal{L}^\dagger_L \mathcal{L}_R:)_C 
\end{align*}
\begin{align*}
    \text{pathway 3-10: } (\mathcal{L}_R \mathcal{L}^\dagger_R \mathcal{L}^\dagger_L \mathcal{L}_L \mathcal{L}^\dagger_L \mathcal{L}_R)_C = &(:\mathcal{L}_R \mathcal{L}^\dagger_R \mathcal{L}^\dagger_L \mathcal{L}_L \mathcal{L}^\dagger_L \mathcal{L}_R:)_C + \frac{1}{2}(:\mathcal{S}_R \mathcal{L}^\dagger_L \mathcal{L}_L \mathcal{L}^\dagger_L \mathcal{L}_R:)_C \\
    &+ \frac{1}{2}(:\mathcal{L}_R \mathcal{L}^\dagger_R \mathcal{S}_L \mathcal{L}^\dagger_L \mathcal{L}_R:)_C + \frac{1}{4} (:\mathcal{S}_R \mathcal{S}_L \mathcal{L}^\dagger_L \mathcal{L}_R:)_C
\end{align*}

\end{document}